\documentclass[a4paper,10pt]{article}
\usepackage[utf8x]{inputenc}
\usepackage{float}
\usepackage{amsmath}
\usepackage{fullpage}
\usepackage{mdframed}
\usepackage{titlesec}
\usepackage{color}
\usepackage[colorlinks,citecolor=red,linkcolor=blue]{hyperref}

\titleformat{\paragraph}
{\normalfont\normalsize\bfseries}{\theparagraph}{1em}{}
\titlespacing*{\paragraph}
{0pt}{3.25ex plus 1ex minus .2ex}{1.5ex plus .2ex}

\usepackage{listings}
\usepackage{amssymb}
\usepackage{epsfig}
\usepackage{mathtools}
\usepackage{slashed}
\usepackage{breqn}
\usepackage{graphicx}
\renewcommand{\thesection}{\arabic{section}}

\def\beq{\begin{equation}}
\def\eeq{\end{equation}}
\def\bea{\begin{eqnarray}}
\def\eea{\end{eqnarray}}
\def\bmat{\begin{pmatrix}}
\def\emat{\end{pmatrix}}

\def\to{\rightarrow}

\title{Extra Neutral Scalars with Vector-like Fermions at the LHC}

\author{Shrihari~Gopalakrishna~\thanks{shri@imsc.res.in}~, 
Tuhin Subhra Mukherjee~\thanks{tuhin@imsc.res.in}~, 
Soumya Sadhukhan~\thanks{soumyasad@imsc.res.in}~, \\  
\small{The Institute of Mathematical Sciences (IMSc),} \\ 
\small{C.I.T Campus, Taramani, Chennai 600113, India.}
}

\begin{document}
\maketitle

\begin{abstract}

Many theories beyond the standard model (BSM) 
contain new CP-odd and CP-even neutral scalars $\phi = \{A,H\}$, and new vector-like fermions ($\psi_{VL}$). 
The couplings of the CP-odd scalar $A$ to two standard model (SM) gauge bosons
cannot occur from renormalizable operators 
in a CP-conserving sector, but can be induced at the quantum loop level.   
We compute these
effective couplings at the 1-loop level induced by the SM fermions and vector-like fermions, present analytical expressions for them, and plot them numerically. 
Using the 8~TeV Large Hadron Collider (LHC) $\gamma\gamma$, $\tau^{+} \tau^{-}$ and $t \bar t$ channel data, 
we derive constraints on the effective couplings of the $\phi$ to standard model gauge bosons and fermions.
We present the gluon-fusion channel cross-sections of the $\phi$ at the 8~and~14~TeV LHC, 
and its branching-ratios into SM fermion and gauge-boson pairs. 
We present our results first model-independently, and then also for some simple models containing 
$\phi$ and $\psi_{VL}$ in the singlet and doublet representations of $SU(2)$.
In the doublet case, we focus on the two-Higgs-doublet (2HDM) Type-II and Type-X models in the alignment limit. 

\end{abstract}

\section{Introduction}

A long series of experiments culminating in the Large Hadron Collider (LHC) discovery of the Higgs boson at a mass of about $125$~GeV has firmly established the standard model~(SM) as the correct description of Nature 
up to an energy scale of a few hundred GeV.
With this discovery, the theoretical puzzle as to why the Higgs boson remains this light when quantum effects should correct it to the highest scales present in the theory 
(such as the Planck scale) comes to the fore. 
This problem of the stability of the electroweak (EW) scale is the well known hierarchy problem of the SM. 
This could be a clue that some new physics beyond the standard model (BSM) is present near the EW scale 
which renders it stable against quantum corrections, making it natural.
Many theoretical proposals have been made for this new physics (for reviews see Ref.~\cite{EWSBrev.BIB}), and they usually contain new particles at the TeV energy scale. 
We are poised at a very interesting time when the LHC is probing this energy scale and can tell us if one of these proposals is realized in Nature. 

Among the possibilities of BSM physics that makes the EW scale natural are models in which the Higgs-doublet of the SM is a pseudo-Nambu-Goldstone boson (pNGB).
Concrete realizations of this idea, for example, are in models of little-Higgs, composite-Higgs and extra dimensions 
(for reviews see Refs.~\cite{LittleHiggsRev.BIB,Contino:2010rs,Davoudiasl:2009cd} respectively). 
In such models, in addition to the CP-even Higgs boson, there could be new CP-odd scalar ($A$) and CP-even scalar ($H$), which we denote collectively as $\phi=\{A,H\}$,
that are also pNGBs due to which their mass is much lower than the cut-off scale.  
Also, new heavy vector-like fermions (VLF, denoted as $\psi_{VL}$) are usually required, that along with the SM fermions (SMF), 
complete some representation of a bigger group containing $SU(2)\otimes U(1)$.
The new vector-like fermions can include vector-like quarks (VLQ) and vector-like leptons (VLL) and may be present 
in addition to the usual SM quarks (SMQ) and leptons (SML).  
By vector-like fermions we mean that fermions in a representation of the SM gauge-group and in its conjugate representation both appear in the theory 
(for more details see for example Ref.~\cite{VLFDir.REF}). 
Some supersymmetric models also include vector-like matter, and thus have $\phi$ and $\psi_{VL}$ both present, along with many superpartners.

The phenomenology of a CP-odd scalar at the LHC can be quite distinct as compared to a CP-even scalar (such as the SM Higgs boson),
and one focus of this work is to elucidate this aspect.
If CP-invariance is not spontaneously broken by an $A$ vacuum expectation value (VEV), i.e. if $\left< A \right> = 0$, as we assume here,  
$AW^+W^-$, $AZZ$ (collectively called $AVV$ couplings), and also $A\gamma\gamma$ and $AZ\gamma$ couplings cannot arise from renormalizable operators. 
Also, the last two do not arise from renormalizable operators because of unbroken electromagnetic (EM) gauge invariance, the same reason why $h\gamma\gamma$ and $h\gamma Z$ 
are zero at the renormalizable level. 
These can then only result from higher-dimensional operators generated at loop-level.
In contrast, for the CP-even SM Higgs boson (denoted as $h$), the $hW^+W^-$ and $hZZ$ couplings are generated at tree-level from dimension-four operators 
after electroweak symmetry breaking (EWSB), i.e. with $\left< h \right> = v/\sqrt{2}$. 
Therefore, generically speaking, the $AW^+W^-$ and $AZZ$ effective couplings, generated at loop-level, are much smaller in magnitude compared to the tree-level $hW^+W^-$ and $hZZ$
couplings; 
the $A \gamma\gamma$ and $h \gamma \gamma$ effective couplings are both loop suppressed and small, and similarly the $A\gamma Z$ and $h\gamma Z$ are also both loop suppressed. 
Thus, similar to the $h$, the $gg\to A$ ``gluon-fusion'' channel is important at the LHC, while compared to the $h$, the vector-boson fusion channel of $A$ is much suppressed.  
The alternate possibility of $\left< A \right> \neq 0$ is not discussed here but is considered for instance in Refs.~\cite{Bento}.  

Turning next to $\phi=\{A,H\}$ couplings to fermions, we include $\phi$ couplings to new vector-like fermions at the tree-level. 
Furthermore, if $\phi$ is part of a doublet, it couples also to SM fermions at the tree-level (similar to $h$).
We consider the case when $\phi$ couples significantly only to third generation SM fermions, a situation common in many BSM extensions. 
Thus, the relevant couplings to SM fermions are $\phi b \bar b$, $\phi \tau^{+}\tau^{-}$ and $\phi t \bar t$. 
If the $\phi b\bar b$ coupling is sizable, $b\bar b \to \phi$, $bg\to b\phi$ and $gg\to b \bar b\phi$ can be important production channels of the $\phi$.
However, we do not include these production channels in this work, but restrict ourselves only to the gluon-fusion channel.

Mostly, we restrict ourselves to the situation when $m_\phi < 2 M_{VL}$ so that $\phi$ cannot decay to a pair of VLFs.
If the $\psi_{VL}$ is light enough they can also be studied directly at the LHC as discussed for instance in Ref.~\cite{VLFDir.REF} and references therein. 
However, if they are too heavy to be directly produced at the LHC, but the $\phi$ (or $h$ as studied in Ref.~\cite{Ellis:2014dza}) can be directly produced and it's couplings measured, 
the VLF contributions to the $\phi$ effective couplings we derive here can be useful in probing the $\psi_{VL}$ indirectly.

We identify the lighter CP-even state ($h$) to be the 125~GeV state discovered, and whose properties measured, at the LHC.
The $h$ couplings measured at the LHC so far largely agree with the SM, at least to about a few tens of percent,
and the magnitude of the $hVV$ coupling (with $V=\{W_\mu^\pm, Z_\mu \}$) is constrained to be close to the SM coupling at the few tens of percent level.
This will be realized in the so called ``decoupling limit''~\cite{Gunion:2002zf}, or more generally in the ``alignment limit''~\cite{Bhattacharyya:2015nca}.
In order to capture many different BSM models, we perform a model-independent effective theory analysis of the $\phi$ coupled to SM fields. 
We present the constraints from the recent $8~$TeV LHC run using the $\gamma\gamma$, $\tau^{+}\tau^{-}$ and $t \bar t$ channels, 
and present the signal cross section (c.s., $\sigma$) at the LHC as a function of the effective-couplings of the $\phi$ (denoted by $\kappa$) and
the branching ratio ($BR$) into these modes.
We do not focus much on the $ZZ$ and $W^+W^-$ decay channels of the $\phi$ as the branching-ratio into these modes are much smaller than the other modes due
to $AVV$ coupling being generated only at the loop-level, and the $HVV$ coupling being zero in the alignment limit. 
We also present many simple models containing $A$ and $\psi_{VL}$ in SU(2)-singlet and doublet representations. 
For $A$ in a doublet, we restrict ourselves to the two-Higgs-doublet model (2HDM) Type-II and Type-X. 
We present the 1-loop analytical expressions for the $\{Agg, A\gamma\gamma, A\gamma Z\}$ effective couplings 
induced by SMFs and VLFs in each of these models;
as a function of the model parameters, we plot numerically these effective couplings and the $BR$ into the 
$\gamma\gamma$, $\gamma Z$
and fermion final states. 
These are some of the main results of this work.

In previous studies, one of us has considered the implications of models with VLQs and 
VLLs coupled to the lighter CP-even Higgs boson $h$ in Ref.~\cite{Ellis:2014dza}, 
and the direct LHC signatures of VLQs in Refs.~\cite{VLFDir.REF};
this work complements them by considering aspects of heavier neutral CP-odd and CP-even scalars $A,H$.
In Ref.~\cite{Gopalakrishna:2015dkt} we study many aspects dealt with in this paper but in a specific little-Higgs model, 
the SU(6)/Sp(6) model by Low, Skiba and Smith~\cite{Low:2002ws}. 
We also list there many little-Higgs models that contain a 2HDM structure. 
The results of this paper are useful in deriving constraints and prospects of such models. 

From the vast literature, we give a sampling below of studies that deal with extra BSM neutral scalars, 
have overlap with our work and that take into account the recent LHC 8~TeV constraints. 
We also mention how our work complements them.
There exist several studies which present $\sigma(pp \to A)$ (see for example Refs.~\cite{Liebler:2015bka,Bagnaschi:2014zla})
in the context of 2HDM, minimal supersymmetric standard model (MSSM) and next-to-MSSM (NMSSM).
We highlight the effects of VLFs on $\sigma (gg \to A)$ in various SM extensions including 2HDM-II and 2HDM-X.
Refs.~\cite{Coleppa:2012eh,Burdman:2011ki} consider the possibility that the observed 125~GeV state at the LHC is a CP-odd scalar, and the former shows that this possibility is disfavored by the LHC data.
Refs.~\cite{Dumont:2014wha,Dumont:2014kna} analyze 2HDM Types~I~and~II taking into account the 125~GeV LHC data, all pre-LHC constraints and results of the heavy-Higgs searches in various channels.
Ref.~\cite{Cheung:2013rva} performs a global fit of general 2HDMs using ATLAS, CMS and Tevatron results.
Ref.~\cite{Coleppa:2013dya,Das:2015qva,Bhattacharyya:2013rya,Cheung:2014oaa,Broggio:2014mna} shows the allowed parameter space of 2HDM-II, applying theoretical (perturbativity, unitarity and vacuum stability) 
and experimental (LEP, Tevatron and LHC 125~GeV Higgs data, precision observables and $B$-physics and electric dipole moment measurements) constraints. 
Ref.~\cite{Chen:2013rba} also includes the heavy Higgs exclusion limits to constrain the 2HDM.
LHC 8~TeV constraints on the 2HDM parameter-space are also discussed in Refs.~\cite{Drozd:2012vf,Grinstein:2013npa,Celis:2013rcs,Chang:2012ve,Eberhardt:2013uba}.
The heavy neutral scalars of the 2HDM, namely $A$ and $H$, are studied in Ref.~\cite{Coleppa:2014hxa}, where the LHC 8~TeV exclusion and 14~TeV reach from the processes 
$gg \to H \to AZ$ and $gg \to A \to HZ$ are presented.
Ref.~\cite{Dev:2014yca} constructed an $SO(5)$ symmetric 2HDM which naturally realizes the "alignment limit"
and puts constraints on it's parameter space from the 8 TeV LHC data.
Ref.~\cite{Baglio:2014nea} puts limits on the the triple Higgs couplings
and presents a set of benchmark points for probing SM-Higgs pair production and the search of heavy Higgs bosons through non-standard decay channels
(i.e decays of $A,~H$ that involves at least one Higgs boson in the final state).
Ref.~\cite{Bernreuther:2010uw} calculates the loop factors for the $AVV$ couplings in the MSSM and the 2HDM with a heavy chiral fourth generation.  
Ref.~\cite{Bernreuther:2009ts} studies $A\rightarrow WW,ZZ$ decays and compares this with the corresponding CP-even scalar decays
in 2HDM-II, and also with a chiral fourth generation or additional heavy vector-like quarks (VLQ) added. 
In addition to these, here we also include the effects of VLFs on $A \to \gamma\gamma,Z\gamma$ decays.
An effective Lagrangian analysis of new heavy scalar particles is presented in Ref.~\cite{deBlas:2014mba}.
Various VLF models and related phenomenological issues are also studied in Refs.~\cite{delAguila}. 
Many of these studies are done with specific models in mind while we present the LHC limits and signal c.s. in a model-independent manner, and using these, 
derive results for the models we introduce, and also for some of the models above.

The paper is organized as follows: 
In Sec.~\ref{ModInd.SEC} we present a model-independent analysis of the CP-odd and CP-even neutral scalars $\phi$, 
present constraints on its effective couplings from the 8~TeV LHC run, 
the LHC gluon-fusion c.s., and $BR$ into SM fermion and gauge boson decay modes. 
In Sec.~\ref{Models.SEC} we present many simple models containing $\phi$ and $\psi_{VL}$ as $SU(2)$ singlets or doublets. 
For each of these models, we work out the 1-loop effective couplings of the $\phi$ and present its $BR$ into two body decay modes.  
One can read-out the current constraints and gluon-fusion c.s of the $\phi$ at the LHC for each of these models in conjunction with the results in Sec.~\ref{ModInd.SEC}. 
The models considered include $\phi$ as an $SU(2)$ singlet, or contained in the 2HDM, with correspondingly the $\psi_{VL}$ also in singlet or doublet representations. 
We offer our conclusions in Sec.~\ref{Concl.SEC}. 
For the various models we discuss, we compile expressions for the mass eigenvalues and mixing angles in App.~\ref{effCoups.APP}, and the 1-loop effective couplings in App.~\ref{kAvvGen.App}.

\section{Model-independent Analysis}
\label{ModInd.SEC}

In this section, we define an effective Lagrangian with couplings of the neutral scalars, 
CP-odd $A$ and CP-even $h, H$ to SM gauge bosons and fermions.
We denote the neutral scalars collectively as $\phi$.
In models that contain two CP-even scalars, we identify the lighter one ($h$) as the $125$~GeV scalar observed at the LHC.
For the heavier states $(A,H)$, we show the constraints from the 8~TeV LHC, 
signal c.s. $\sigma \times$ $BR$ into various SM two body final states at the $8$~and~$14$~TeV LHC, 
as a function of the effective couplings and $m_{\phi}$. 
For any given new physics model, one can obtain this effective Lagrangian by integrating out heavier fields, following which  
the results of this section can then be used to obtain the LHC limits and gluon-fusion cross-section in that model.

CP invariance requires the CP-odd scalar $A$ coupling to SM gauge bosons to be \emph{only} via higher dimensional operators. 
The CP-even scalars can couple to the massive gauge bosons at tree level. 
Showing only the new physics terms, the effective Lagrangian for any neutral scalar $\phi$ is  
\begin{align}
\label{LEff.EQ}
\mathcal{L}_{eff} =  
      &  \frac{1}{2} \partial_{\mu}\phi  \partial^{\mu} \phi  - \frac{1}{2} m_{\phi}^{2} \phi^{2} - y_{\phi f_{i}f_{i}} \phi \bar f_{i} X f_{i} \\
      &+ y_{\phi WW} \phi W^{\mu} W_{\mu} + y_{\phi ZZ} \phi Z^{\mu} Z_{\mu}  - \frac{1}{64 \pi^{2} M}\kappa_{\phi \gamma \gamma}\phi Y_{\mu\nu\sigma\tau} F^{\sigma\tau} F^{\mu \nu}- \frac{1}{32 \pi^{2} M}\kappa_{\phi \gamma Z} \phi Y_{\mu\nu\sigma\tau} F^{\sigma\tau} Z^{\mu \nu} \nonumber \\ 
      &-\frac{1}{64 \pi^{2}M} \kappa_{\phi gg} \phi Y_{\mu\nu\sigma\tau} G^{\sigma\tau} G^{\mu \nu} -\frac{1}{64 \pi^{2} M}\kappa_{\phi ZZ} \phi Y_{\mu\nu\sigma\tau} Z^{\sigma\tau} Z^{\mu \nu} -\frac{1}{32 \pi^{2} M} \kappa_{\phi W W}\phi  Y_{\mu\nu\sigma\tau} W^{\sigma\tau} W^{\mu \nu}, \nonumber
\end{align}
where $X= \gamma_{5}, Y_{\mu\nu\sigma\tau}= \epsilon_{\mu\nu\sigma\tau}$ for the CP-odd scalar,
while $X= I$ (identity matrix), $Y_{\mu\nu\sigma\tau}= g_{\mu\sigma} g_{\nu\tau}$ for the CP-even scalar.
Here $\kappa_{\phi ij}$~s contain other fermion and gauge boson loop contributions. 
Tree level scalar gauge boson couplings $y_{\phi ZZ}, y_{\phi WW}$ are zero for the $A$.
We have defined the dimensionless effective couplings $\kappa$ by pulling out a new-physics mass-scale $M$ in the effective $\phi VV$ terms. 
For the numerical results we show, we set $M = 1$~TeV from now on and show only $\kappa$, and for other values of $M$, the $\kappa$ can easily be rescaled. 
Although we have defined the effective couplings $\kappa$ by extracting a heavy new-physics mass scale $M$, SM fermion contributions are to be included when present. 
Eq.~(\ref{LEff.EQ}) is an effective Lagrangian at a scale just above $m_{\phi}$. Heavy BSM fermion and the SM fermion contributions are to be included in $\kappa$ before comparing with the plots we show 
in this section.
For various simple SM extensions detailed in Sec.~\ref{Models.SEC} we compute the $\kappa$'s and present them in App.~\ref{effCoups.APP}. 
If SM fermions contribute and can go onshell, the $\kappa$ are complex.
In this case, the $\kappa_{\phi VV}$ that appear in our plots in this section should be read as $|\kappa_{\phi VV}|$.
We assume $y_{\phi f_i f_i}$ to be real in this work. 

The CP-odd scalar can decay to SM gauge bosons or fermions. In terms of the $\kappa$ and $y$'s defined above, the decay rates to different final states are 
\begin{align}
&\Gamma(\phi \rightarrow Z \gamma) = \frac{1}{32 \pi} \left(\frac{\kappa_{\phi Z\gamma}}{16 \pi^{2} M}\right)^{2}m_{\phi}^{3} (1-r_{Z})^{3}, \hspace{1cm}\Gamma(\phi \rightarrow g g) = \frac{1}{8 \pi} \left(\frac{\kappa_{\phi  g g }}{16 \pi^{2} M}\right)^{2}m_{\phi}^{3}, \nonumber \\
&\Gamma(\phi \rightarrow ff) = \frac{N_c}{8 \pi} y_{\phi ff}^{2} m_{\phi}(1- 4 r_{f})^{n/2}, \hspace{2.2cm}\Gamma(\phi \rightarrow \gamma \gamma) = \frac{1}{64 \pi }\left(\frac{\kappa_{\phi \gamma \gamma}}{16 \pi^{2} M}\right)^{2} m_{\phi}^{3},
\label{GmA2XX.EQ}
\end{align}
where $n=3$ and $n=1$ for CP-even and CP-odd scalars respectively, $r_f = m_f^{2}/m_\phi^{2}$, $r_Z = m_Z^{2}/m_\phi^2$
with $N_c=3$ for quarks and $1$ for leptons.
Here we have defined $\Gamma(\phi \to gg)$ to have an extra factor of 8 compared to $\Gamma(\phi \to \gamma\gamma)$
anticipating a color factor. It turns out however that for a quark in the loop, the color factor in the $\Gamma(\phi \to gg)$
is actually 2. This will get compensated for in $\kappa_{\phi gg}$ (see for example Eq.~(\ref{kphiag})).
Using these expressions, one can work out the $BR$ of the $\phi$ into these final states in any new physics model.

We turn next to discussing limits from the $8$~TeV LHC and the gluon-fusion cross-section at $14$~TeV. 
To obtain the limits on the effective couplings $\kappa$ and $y$, we use upper-limits ($UL$) from recent LHC analysis on 
$\sigma(pp \rightarrow \phi) \times $ BR $(\phi\rightarrow XX)$, and the currently relevant constraints are $XX = \{\gamma\gamma, \tau^{+} \tau^{-}, t \bar t\}$.  
We take the limits on the $\gamma\gamma$ channel from the CMS analysis Ref.~\cite{CMS:2014onr} which has an upper limit up to $M_\phi$ of 850 GeV,
on the $\tau^{+} \tau^{-}$ channel from the ATLAS analysis Ref.~\cite{atlas_ichep} up to $M_\phi$ of 1000 GeV,  
and from the ATLAS analysis Ref.~\cite{madalinatt} for the $t \bar t$ channel.    
Using these we constrain the effective couplings of Eq.~(\ref{LEff.EQ}).  

At the LHC, the $\phi$ can be produced by $gg\to \phi$ (called gluon-fusion channel), which starts at the 1-loop level when $\phi$ couples to colored fermions. 
In addition to the above production channel, if $\phi$ couples to $b$-quarks, there are additional production channels,  
namely, $b \bar b \to \phi$ (called $b\bar b$-fusion), $bg\to b \phi$ and $gg\to b \bar b\phi$ (called $b$-quark associated production) channels;
how these compare with the gluon-fusion channel depends on how large the $b \bar b \phi$ coupling is in a given model.
For instance, for $y_{b\phi}=0.5$, we find that the production rate via $b\bar b$-fusion and $b$-quark associated production channels becomes comparable to the gluon-fusion channel with $\kappa_{\phi gg} \approx 20$. 
We include only the gluon-fusion channel in this study, but in models with a large $b \bar b\phi$ coupling, the $b\bar b$ fusion and $b$-quark associated production channels
may have to be included, which we do not do here.
For a study involving the $b$-quark associated production channels of the $h$ including $gg\to b\bar b h$, see Ref.~\cite{Maltoni:2003pn}.
One can separately study the $b$-quark associated production channels by tagging on the final state $b$-jet as discussed in Ref.~\cite{atlas_ichep}.
Ref.~\cite{Kozaczuk:2015bea} has recently studied $b \bar b$~fusion and $b$-quark associated production channels for a light CP-odd scalar.
Although there are some LHC limits using $b$-tagged events to which the $b\bar b$ decay mode and the $b$-quark associated production channels contribute,
we do not include them in our analysis here. 
So far these results have been presented for $m_{\phi} < 350$~GeV (see Refs.~\cite{Chatrchyan:2013qga,Aad:2014xzb,CMS:2014nm}).

Rather than compute the $A, H$ production rate at the LHC ourselves, we relate it to the SM Higgs production rate at the same mass, 
and make use of the vast literature on $h$ production rate.
Since $\sigma(gg\to \phi) \propto \Gamma(\phi \to gg)$, we can write the $\sigma * BR$ for $\phi$ production followed by decay into the final-state $XX$ as  
\begin{eqnarray}
\sigma(gg\to \phi) = \frac{\Gamma(\phi \rightarrow gg)}{\Gamma (h \rightarrow gg)} \times \sigma(g g \rightarrow h) \ .   
\label{ggAggh.EQ}
\end{eqnarray}
From Eq.~(\ref{GmA2XX.EQ}) we compute $\Gamma(\phi \to gg)$ for a given $\kappa_{\phi gg}$ 
and using Eq.~(\ref{ggAggh.EQ}) we apply the upper-limit ($UL$) from the $8$~TeV LHC quoted above for various $BR(\phi\to XX)$. 
For our numerical work, we calculate $\Gamma(h \to gg)$ in the SM and take $\sigma(g g \rightarrow h)$ from Ref.~\cite{Baglio:2010ae}. 
We assume here that the dependence on the PDF, and the acceptance at the LHC for $A, H$ and $h$ are not very different, which should be reasonable assumptions. 
For the decay $A\to XX$, the final-states $XX$ we consider are $\gamma\gamma$, $\tau^{+}\tau^{-}$ and $t \bar t$ as these are currently the significant ones.
We compute the $BR(A\to XX)$ using Eq.~(\ref{GmA2XX.EQ}).   
If $A, H$ are fairly close in mass, i.e. closer than the experimental resolution to separate them (say $30\,\%$ of $m_{\phi}$), 
and no kinematic variables can separate them, we should include all of them into the $\sigma * BR$ above. 

\begin{figure}[]
\centering
{\label{kgg_prod_8TeV}}\includegraphics[width=0.32\textwidth]{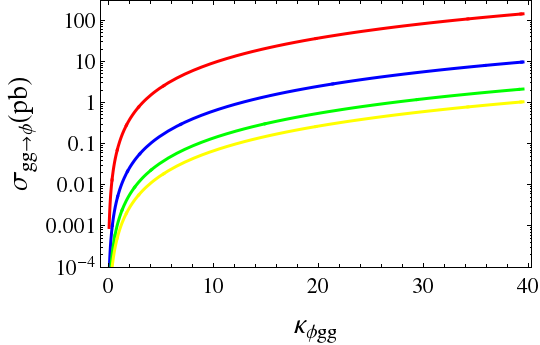}\hspace{1em}%
{\label{kgg_prod_14TeV}}\includegraphics[width=0.32\textwidth]{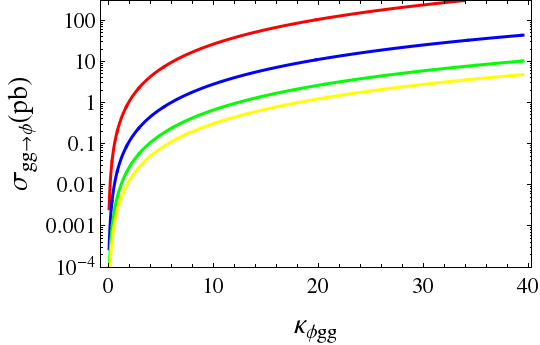}
\caption{$\sigma(gg \rightarrow \phi)$ (in pb) at the 8~TeV LHC (left) and 14~TeV LHC (right) for $m_{\phi}=200$~GeV (red), $500$~GeV (blue), $800$~GeV (green) and $1000$ GeV (yellow).}
\label{kgg_prod}
\end{figure}

\begin{figure}[]
\centering
{\label{kgg_gamma_8TeV}}\includegraphics[width=0.33\textwidth]{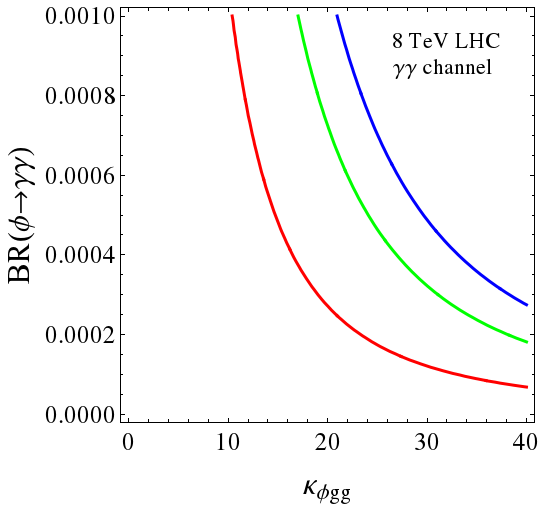}\hspace{1em}%
{\label{kgg_tau_8TeV}}\includegraphics[width=0.31\textwidth]{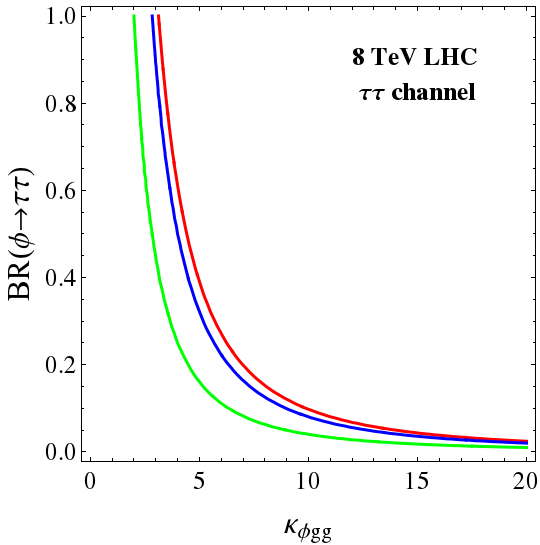}
{\label{kgg_yt_TeV}}\includegraphics[width=0.31\textwidth]{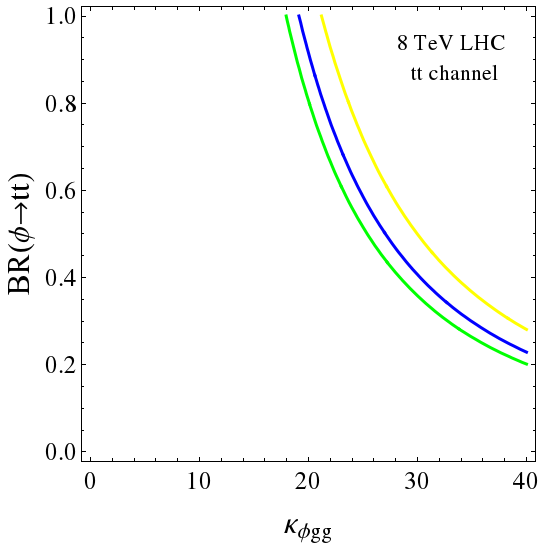}
\caption{8~TeV LHC constraints from the $\gamma\gamma$ channel (left), $\tau^{+}\tau^{-}$ channel (middle) and $t\bar t$ channel (right), 
for $m_{\phi}=200$~GeV (red), $500$~GeV (green), $800$~GeV (blue) and $1000$ GeV (yellow). 
The regions to the top and right of the curves are excluded at the 95\,\% CL level.}
\label{kgg_8TeV}
\end{figure}

\begin{figure}[]
\centering
{\label{kgg_gamma_200_BR}}\includegraphics[width=2.0in]{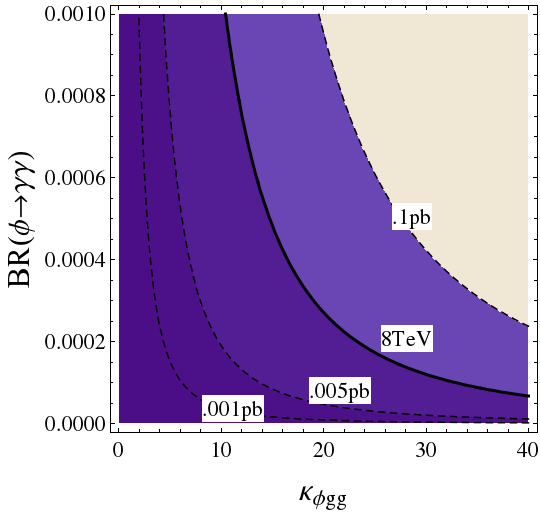}
{\label{kgg_gamma_500_BR}}\includegraphics[width=2.0in]{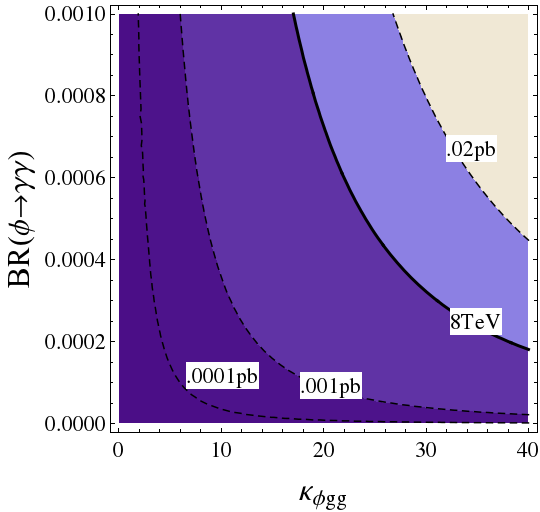}
{\label{kgg_gamma_800_BR}}\includegraphics[width=2.0in]{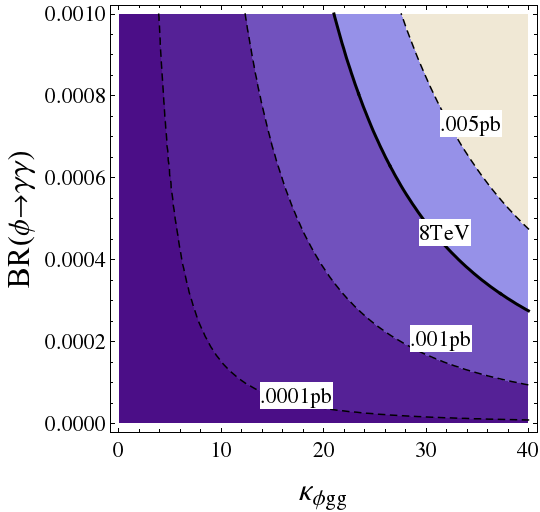}\\
{\label{kgg_tau_200_BR}}\includegraphics[width=1.92in]{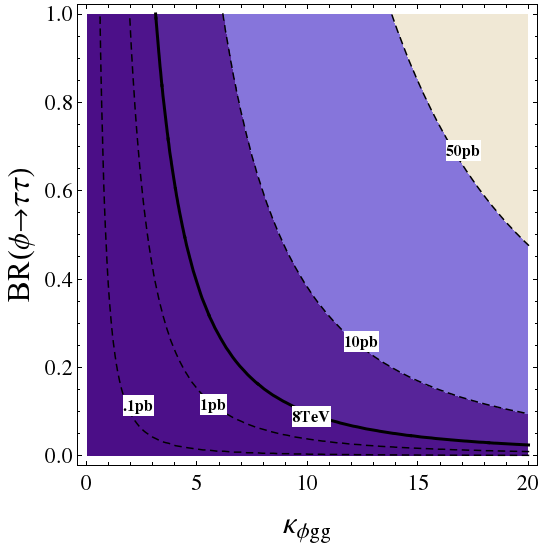}\hspace{1em}%
{\label{kgg_tau_500_BR}}\includegraphics[width=1.92in]{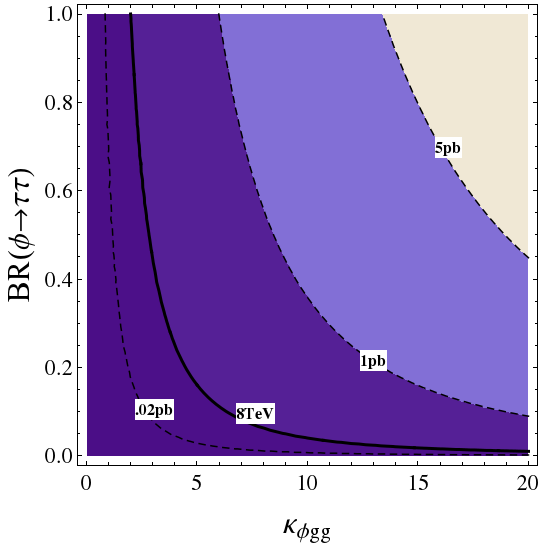}\hspace{1em}%
{\label{kgg_tau_800_BR}}\includegraphics[width=1.92in]{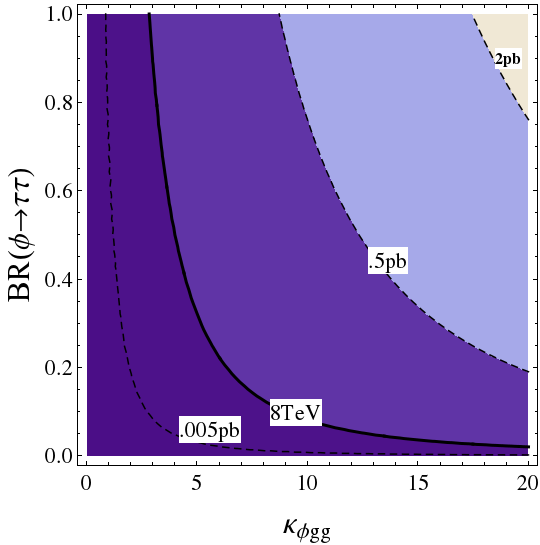}\\
{\label{kgg_tt_500_BR}}\includegraphics[width=1.92in]{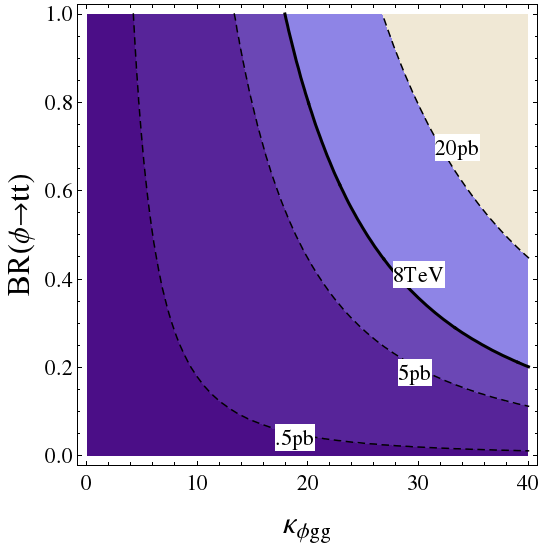}\hspace{1em}
{\label{kgg_tt_800_BR}}\includegraphics[width=1.92in]{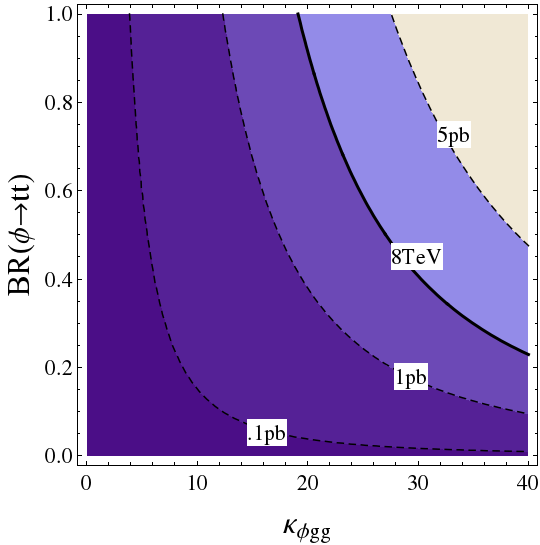}\hspace{1em}%
{\label{kgg_tt_1000_BR}}\includegraphics[width=1.92in]{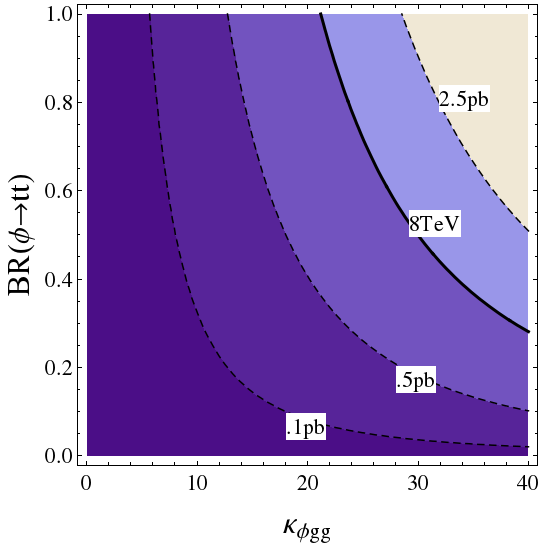}
\caption{Contours of the 14~TeV LHC $\sigma \times BR$ (in $pb$) in the $\gamma\gamma$ channel (upper-row)
and the $\tau^{+}\tau^{+}$ channel (middle-row) for $m_{\phi}=200$~GeV (left), 500~GeV (middle), 800~GeV (right), and in the $t\bar{t}$ channel (bottom-row) for $m_A=500$~GeV (left), 800~GeV (middle), 1000~GeV (right). 
The region to the right of the contour labeled '8~TeV' is excluded at the $95\,\%$ CL level from 8~TeV LHC result.}
\label{kgg_BR}
\end{figure}

In Fig.~\ref{kgg_prod} we show $\sigma(gg \to \phi )$ at the 8~TeV LHC (left plot) and 14~TeV LHC (right plot) as a function of $\kappa_{\phi gg}$.
$\sigma(gg\to \phi)$ is obtained using Eq.~(\ref{ggAggh.EQ}) and the $\sigma(gg\to h)$ from Ref.~\cite{Baglio:2010ae} as mentioned earlier. 
In a given new physics model, one can compute $\kappa_{\phi gg}$ and then use these plots to obtain the $\sigma(gg\to \phi)$. 
Using the $\sigma(gg\to \phi)$, we obtain constraints from the 8~TeV LHC data as a function of the BR into a particular mode. 
We show this in Fig.~\ref{kgg_8TeV} obtained from the $\gamma\gamma$, $\tau^{+}\tau^{-}$ and $t\bar t$ channels. 
The regions to the top and right of the curves are excluded at the 95\,\% CL level.
In the $\gamma\gamma$ channel, the bound is strongest for $m_{\phi}=200$~GeV since the experimental exclusion is tightest at that mass. 
We see that there is no constraint from this channel for $BR(\phi \to \gamma\gamma) \lesssim 10^{-4}$ for the range of $\kappa_{\phi gg}$ shown.  
From the $\tau^{+}\tau^{-}$ channel, we find the strongest limit for $m_\phi$ of about 500~GeV since the experimental exclusion is tightest at that mass. 
We show in Fig.~\ref{kgg_BR} the total $\sigma(gg\to \phi) * BR(\phi \to XX)$ contours (in $pb$)
for $XX = \{ \gamma\gamma, \tau^{+}\tau^{-}, t \bar t \}$ at the 14~TeV LHC,
making use of the fact that the total $\sigma(gg \to \phi \to XX) \propto \kappa^2_{\phi gg}*BR(\phi \to XX)$, omitting kinematic factors independent of couplings.
Thus, each mode $XX$ can be considered and presented independently of the others as we do here.
The $95\,\%$ CL LHC exclusion discussed above is also shown labeled as '8~TeV'. 
If the $\phi bb$ coupling is large, i.e. bigger than about $0.5$, inclusion of the $b$-fusion and $b$-associated production channels 
(along with the $\phi gg$ channel that we have included here) 
could result in a stronger exclusion than we obtain here.  

As already mentioned, the model-independent results presented in this section can be used to obtain the LHC constraints and gluon-fusion c.s.~in any particular model by 
computing first the effective couplings in that model. 
We compute the effective couplings in many simple models next.

\section{Models}
\label{Models.SEC}
In this section we consider some specific models for the neutral CP-odd and CP-even scalars $A,H$ 
and study their LHC production and decays into 2-body final states.
We compute the decay rates assuming a sharp turn-on at threshold of the 2-body final state.   
The goal is to capture in simple models many of the features present in realistic BSM models as far as the LHC phenomenology 
of $A,H$ is concerned.
As before, we collectively denote $A,H$ as $\phi$.
We mostly focus on the situation when $m_\phi < 2 M_{VL}$ and do not focus on the phenomenology due to the $\phi$ decaying to a pair of on-shell VLF.   
We first consider the models where $\phi$ is an $SU(2)$ singlet and couples to $SU(2)$ singlet VLF 
(singlet $A$ with a vector-like up-type-singlet or $SVU$~model, or with a down-type-singlet or $SVD$ model) 
and $SU(2)$ doublet VLF (singlet $A$ with minimal vector-like quark doublet or $SVQ$~model).
We next consider effective models with $\phi$ in an $SU(2)$ doublet, 
with the two SU(2) doublet scalars $\Phi_1$ and $\Phi_2$ both having hypercharge $+1/2$.
The 2HDMs we consider are either Type-II like or Type-X like. 
We notate the Type-II like models, for example, as 
$MVQD$ for minimal vector-like extension with VLQ doublet $Q$ and down-type VLQ singlet $D$, 
and $MVQU$ for a similar model with an up-type VLQ singlet $U$ instead,
and a similar model with the 2HDM Type-X structure instead as $MVQDX$. 
We include subscripts depending on which Higgs doublets the fermions couple to, i.e. 
$MVQD_{ij}$ will mean that the model has one VL-quark doublet $\psi$, and one down-type VL-quark singlet $\chi$, 
with the couplings $\bar \psi_L \chi_R \Phi_i$ and $\bar \psi_R \chi_L \Phi_j$ turned on. 
Among our example models are some that mimic BSM models that have $\phi$ Yukawa couplings with an SMQ and a VLQ, 
for example, the 3rd generation $SMQ$ with an up-type singlet-VLQ to give the $MVU$ model. 
 
Many of the effects we present are similar for the CP-odd and CP-even scalars $A,H$. 
One important difference between the $A$ and $H$ is that at tree-level, 
the $AVV$ (with $V=\{W,Z\}$) couplings are zero and are only generated by SM and BSM fermions at the loop level, 
while the $HVV$ couplings could be nonzero at tree-level.
However, in the alignment limit we consider (discussed later), the $HVV$ couplings are zero. 
Although the $\Gamma(H \to \gamma \gamma)$ gets a contribution from the charged scalar ($H^{\pm}$) loop, 
while $\Gamma(A \to \gamma \gamma)$ does not, this is very small compared to the fermionic contributions~\cite{Djouadi:2005gj}.
Thus in the alignment limit the $A$ and $H$ have very similar phenomenology. 
Therefore we will mostly present the phenomenology of the CP-odd scalar $A$, 
and where relevant, we will also contrast it with the situation for the $H$.
Since the tree-level $HVV$ coupling is zero in the alignment limit, in the appendix we only give 
the expressions for the fermion contributions to the $\kappa_{\phi VV}$. 
For the SM Higgs we must include the $W$ loop contribution to
$\kappa_{h \gamma \gamma}$ and $\kappa_{h Z \gamma}$ which we do not present here.

\subsection{Model with an $SU(2)$ singlet $A$ with VLQ-VLQ Yukawa couplings}
We start by considering some models with an $SU(2)$ singlet $A$ coupled to $SU(2)$ singlet or doublet VLFs. 
For an SU(2) singlet $\phi$ one cannot write Yukawa couplings with chiral SMFs, 
and thus $gg\phi$ and $\gamma\gamma\phi$ couplings
can only be induced by VLFs, if they are present, as we explicitly show here. 
Thus, LHC signals of a BSM singlet $\phi$ becomes possible if colored VLFs are coupled to it. 

\medskip
\noindent \underline{\bf $SVU$ model}: 
We study a model, which we call $SVU$-model, with an $SU(2)$ singlet CP-odd scalar $A$, 
coupled to an $SU(2)$ singlet, $SU(3)$ triplet VLQ ($\psi$)  
with hypercharge $Y_{\psi}$.\footnote{A model with only a vector-like lepton singlet is uninteresting for 
$A$ phenomenology since no LHC production channels are significant 
(note that the $Ab \bar b$ coupling is also not possible in this case).} 
Clearly, the electromagnetic charge $Q = Y_\psi$. 
To the SM Lagrangian we add 
\begin{align}\label{model1a}
\mathcal{L} \supset  &\frac{1}{2} \partial_{\mu}A \partial^{\mu}A -\frac {1}{2} m_{A}^{2} A^{2}+ \bar \psi i \slashed \partial \psi + eQ A_{\mu}\bar \psi \gamma^{\mu} \psi -  g Q  \frac{s_{W}^{2}}{c_{W}}Z_{\mu}\bar \psi \gamma^{\mu}\psi \\ \nonumber
                        & + \bar \psi i \slashed D \psi - iy_{A} A \bar \psi \gamma_{5} \psi- m_{\psi} \bar \psi \psi -\frac{\lambda_{A}}{6} A^{2} H^{\dagger} H .
\end{align}
The SM Higgs doublet is written as H here.
Here we have not considered possible terms coupling the $A$ to a SM fermion and a VLF for $Y_{\psi}=2/3,-1/3$
such as $\bar \psi_{L} A u_{R}$, $\bar \psi_{L} A d_{R}$, $\bar q H \psi_{R}$. 
We study this possibility of off-diagonal couplings between the 3rd generation SMQ and a VLQ 
in the context of the SU(2) doublet $\Phi$ in Sec.~\ref{VLF.mix}.

\begin{figure}[]
\centering
{}{\includegraphics[width=0.32\textwidth]{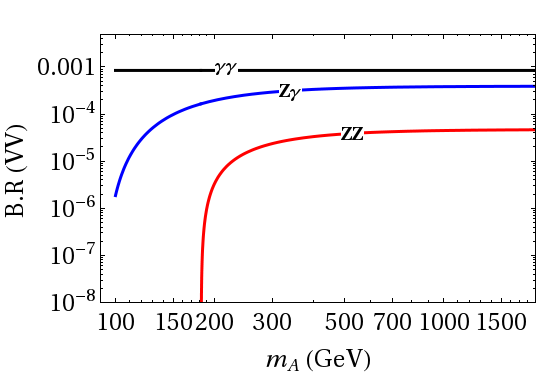}}\hspace{1em}%
{\includegraphics[width=0.32\textwidth]{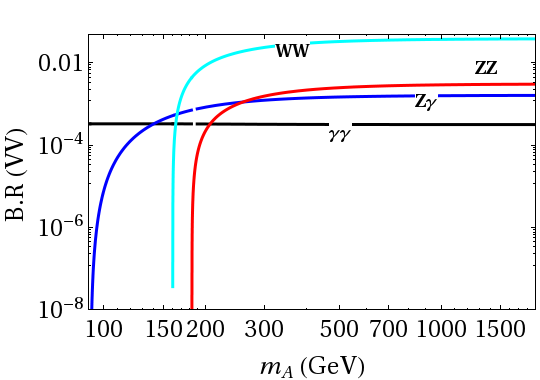}}\hspace{1em}%
\caption{BR $(A\rightarrow \gamma \gamma)$ (black), BR ($A\rightarrow\gamma Z$) (blue), BR($A\rightarrow ZZ$) (red), BR($A\rightarrow WW$) (cyan) as a function of $m_{A}$ with $y_{A}=0.1$ and $m_{\psi}=1000$ GeV for $SVU$ (left) and $SVQ$ (right) models.}
\label{br1a_1b}
\end{figure}

We restrict ourselves to $m_A < 2 M_{VL}$, so that $A$ cannot decay to a VLF pair. 
The possible decay modes of $A$ are to $gg$, $\gamma \gamma$, $Z \gamma$ and $ZZ$ through a VLF loop, 
but no decay to $W^+ W^-$. 
$A$ cannot decay to a pair of SM fermions since such couplings are forbidden by gauge invariance. 
The effective $AV^{\mu}V^{\nu}$ couplings induced by VLFs are given in App.~\ref{kAvvGen.App}.
From these we compute the partial widths and the BR into the above modes.  
In Fig.~\ref{br1a_1b} we plot BR($A \rightarrow \gamma \gamma$), BR$(A \rightarrow Z \gamma)$ and BR$(A\rightarrow ZZ)$ where we chose $Y_{\psi} = 2/3$ as an example. 
BR$(A\rightarrow g g)$ is almost constant at around 0.999.

\begin{figure}[]
\centering
{}{\includegraphics[width=0.32\textwidth]{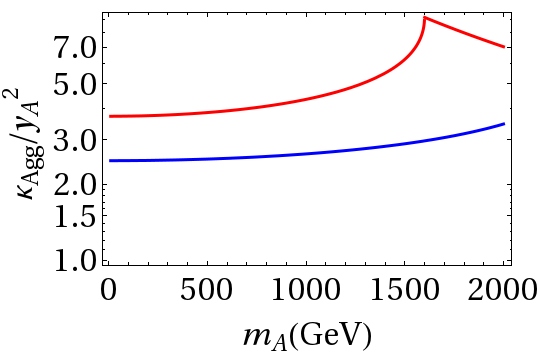}}\hspace{1em}%
\caption{$\kappa_{A gg}/y_{A}^{2}$ as a function of $m_{A}$ for $m_{\psi}$ = 800~GeV (red) and 1200 GeV (blue) for $SVU$~model.}
\label{effcoup_1a}
\end{figure}

In Fig.~\ref{effcoup_1a} we plot $\kappa_{Agg}/y_{A}^{2}$ as a function of $m_{A}$.  
From this, one can read-off the $\sigma(gg\rightarrow A)$ at the 8~and~14~TeV LHC from Fig.~\ref{kgg_prod} in Sec.~\ref{ModInd.SEC}. 
The peaks in Fig.~\ref{effcoup_1a} are due to the VLFs going onshell, although as mentioned earlier, we do not explore its consequences in this work.
In this model, the gluon-fusion c.s. of $A$ is induced only through loops of the heavy VLFs due to which the 8~TeV LHC exclusion limits on 
$\sigma \times BR$ into the $ZZ$ channel (see Ref.~\cite{Chatrchyan:2013yoa}) or the $\gamma \gamma$ channel (see Ref.~\cite{CMS:2014onr}) 
are rather weak, unless $y_A$ becomes so large that perturbativity is lost.

\begin{figure}[h]
\centering
{}{\includegraphics[width=0.32\textwidth]{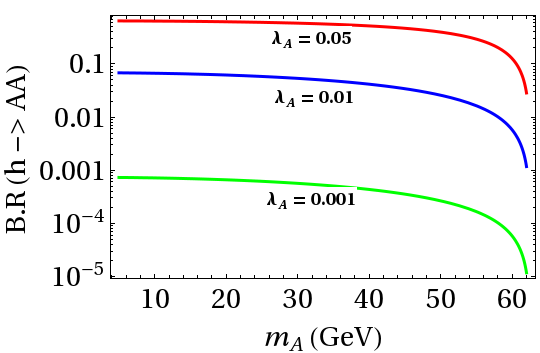}}\hspace{1em}%
{}{\includegraphics[width=0.32\textwidth]{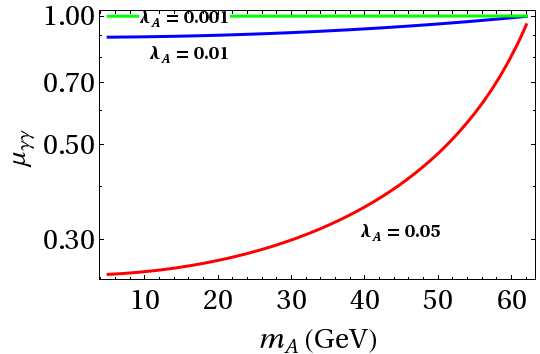}}\\
\caption{  BR $(h\rightarrow AA)$ (left) and $\mu_{\gamma \gamma}$ (right) as a function of $m_{A}$ for $SVU$~model.}
 \label{brmuhaa1a}
\end{figure}

If $m_{A} < m_{h}/2$ (where $h$ is the $125$ GeV Higgs), then $h\rightarrow A A$ becomes kinematically allowed and becomes a means of producing $A$ in addition to the gluon-fusion channel discussed above. 
In Fig.~\ref{brmuhaa1a} we plot BR($h\rightarrow AA$) for $\lambda_{A}$ $= 0.1,~0.05$ and $0.001$.
When this decay is allowed, it will contribute to the Higgs total width thereby modifying the BRs into the other channels.
In particular, it will modify the signal strength $\mu_{\gamma \gamma} = \Gamma(h\rightarrow \gamma \gamma)/\Gamma_{SM}(h\rightarrow \gamma \gamma)$, which is measured to about $10\,\%$ precision 
(see for example Ref.~\cite{ATLAS:2012znl}).
We plot $\mu_{\gamma \gamma}$ in Fig.~\ref{brmuhaa1a}. 
We thus see that the constraint on $\lambda_{A}$ from the 8~TeV LHC is of the order of $0.01$ if $m_{A} < m_{h}/2$.  

\medskip
\noindent \underline{\bf $SVQ$ model}:
We consider a BSM extension, which we call the $SVQ$ model, 
with an $SU(2)$ singlet $A$, and one $SU(2)$ doublet vector-like fermion 
$\psi=\psi_{L,R} = ( \psi_{1L,R} ,~ \psi_{2L,R} )^T $ with hypercharge $Y_{\psi}$. 
To the SM Lagrangian we add
\begin{align}\label{model1b}
\mathcal{L} \supset &\frac{1}{2} \partial_{\mu}A \partial^{\mu}A -\frac {1}{2} m_{A}^{2} A^{2}+ \bar \psi i \slashed D \psi     
                         - iy_{A} A\bar \psi \gamma_{5} \psi -  m_{\psi} \bar \psi \psi - \frac {\lambda_{1}}{4!} A^{4}- \frac{\lambda_{A}}{6}A^{2} H^{\dagger} H \ ,
\end{align}
where the gauge interactions of the $\psi$ are understood and are not explicitly shown.
For $Y_{\psi} =1/6 $ one can add the terms 
$ y'_{u}\bar \psi_{L} \tilde{H} u_{R}+y'_{d}\bar \psi_{L} H d_{R}+iy_{2A} A\bar q_{L} \psi_{R} +h.c $\footnote{We use the notation $\tilde{H} = i \sigma^2 H^*$.}, 
which we will not consider here 
but later in Sec.~\ref{VLF.mix}. 
As in the $SVU$~model, there are no decays to a pair of SM fermions, but unlike there, 
in this model $A\rightarrow W^{+} W^{-}$ decay is also possible through the VLF loop, 
in addition to $gg$, $\gamma \gamma$, $Z \gamma$ and $ZZ$ modes. 
The expressions for
the effective couplings of the $A$ to two SM gauge-bosons
are given in App.~\ref{kAvvGen.App}.
We take $Y_{\psi} =1/6 $ as an example. 

In Fig.~\ref{br1a_1b} we plot the $BR$ of $A$ into $\gamma \gamma$, $Z \gamma$, $ZZ$ and $W^+ W^-$ modes. 
As in $SVU$~model, the $BR$ into $gg$ remains almost constant at around $0.99$ for $m_A \gtrsim 300$~GeV. 
As the $\psi_{1} \psi_{2} W$ coupling ($g$) is greater than the $\psi_{i} \psi_{i} Z$ couplings 
$(g/c_{W})(T_{3} - Q s_{W}^{2})$,  the $BR$ into $W W$ is larger than into $ZZ$.
Again, for the same reasons explained in the $SVU$~model, the exclusion limits from the 8~TeV LHC in the $\gamma\gamma,ZZ,WW$ channels are rather weak in this model also.

The $\sigma(gg\rightarrow A)$ in this model is twice of what was obtained in the $SVU$~model because there are 
two degenerate VLFs in the loop.
The VLFs are degenerate since no Yukawa terms involving the SM Higgs can be written down that can split the masses after
EWSB. 
Since no couplings to a pair of SM fermions exist, there are no $b$-quark initiated production processes possible.

\subsection{Models with $A,H$ in SU(2) doublets (2HDM)}
In the 2HDM we have two scalar doublets, $\Phi_{1}$ and $\Phi_{2}$, which we take to have 
hypercharge $+1/2$.
The physical neutral states are two CP-even scalars ($h$ and $H$) and a CP-odd scalar ($A$).
The Higgs Lagrangian is given by
\beq
  \label{2hdmPhi}
 \mathcal{L}\supset \left| D_{\mu}\Phi_{1} \right|^{2} + \left| D_{\mu}\Phi_{2} \right|^{2} - V(\Phi) \ ,
\eeq
where
\begin{align}
V(\Phi_{1}, \Phi_{2})= &m_{11}^{2} \Phi_{1}^{\dagger}\Phi_{1} + m_{22}^{2} \Phi_{2}^{\dagger}\Phi_{2}-m_{12}^{2}(\Phi_{1}^{\dagger} \Phi_{2}+ h.c) + \lambda_{1}(\Phi_{1}^{\dagger} \Phi_{1})^{2} +\lambda_{2} (\Phi_{2}^{\dagger} \Phi_{2})^{2}  +  \nonumber \\  
                      &\lambda_{3} (\Phi_{1}^{\dagger} \Phi_{1}) (\Phi_{2}^{\dagger} \Phi_{2}) 
         + \lambda_{4} (\Phi_{1}^{\dagger} \Phi_{2}) (\Phi_{2}^{\dagger} \Phi_{1}) +  \frac{\lambda_{5}}{2} [(\Phi_{1}^{\dagger} \Phi_{2})^{2} + h.c.]  \ .
\end{align}
In the limit when $m_{12}^2=0$, the Lagrangian has a discrete $Z_{2}$ symmetry under which $\Phi_{1}\rightarrow -\Phi_{1}$, $d_{R} \rightarrow -d_{R}$ (with all other fields unchanged), 
if the down-type right-handed fermions couple only to the $\Phi_{1}$ and the up-type right-handed fermions only couple to the $\Phi_{2}$ so that there are no tree-level FCNCs 
(see for example Ref.~\cite{Glashow:1976nt}). Nonzero $m_{12}^2$ softly breaks this $Z_{2}$ symmetry. 
We will not consider the hard $Z_{2}$ breaking terms 
$(\Phi_{1}^{\dagger} \Phi_{1} \Phi_{1}^{\dagger} \Phi_{2} + \Phi_{2}^{\dagger} \Phi_{2} \Phi_{2}^{\dagger} \Phi_{1} + h.c)$\footnote{This is a natural choice since if these terms are zero to start with they will not be 
induced at the loop level even if the soft breaking terms are present.}. 
There are eight free parameters in $V$. 
After we fix the minimum of the potential at
$\langle \Phi_1 \rangle =(0,~v_1/\sqrt{2})^T$ and $\langle \Phi_2 \rangle = (0,~v_2/\sqrt{2})^T$, with the constraint $v_1^2+ v_2^2=v^2=(246~{\rm GeV})^2$, the number of free parameters reduces to seven which we take to be 
$m_{A}$, $m_{h}$, $m_{H}$, $m_{H^{\pm}}$, $\tan \beta$, $\alpha$ and $m_{12}^{2}$, 
in a notation that is common in the literature (see Ref.~\cite{Branco:2011iw}).
We parametrize the scalar doublets as
\begin{eqnarray}
 \Phi_{i} = \begin{pmatrix} \phi^{+}_{i} \\ \frac{1}{\sqrt{2}}(v_{i}+ \rho_{i} + i\eta_{i})  \end{pmatrix}, 
\end{eqnarray}
with $v_{1} = v\cos \beta$, $v_{2} = v\sin \beta$ and $\tan \beta = v_2/v_1$.
The physical mass eigenstates are:
a heavy CP-even scalar $H= \rho_{1} \cos \alpha +\rho_{2} \sin \alpha$,
a light CP-even scalar $h = -\rho_{1} \sin \alpha + \rho_{2} \cos \alpha$,
a CP-odd scalar $A = - \eta_{1} \sin \beta + \eta_{2} \cos \beta$, and 
charged scalars $H^{\pm} = - \phi_{1}^{\pm} \sin \beta  + \phi_{2} ^{\pm} \cos \beta$.
All the effective couplings, relevant BRs and the cross sections in the 2HDM can be found in Refs.~\cite{Branco:2011iw,Djouadi:2005gj}.
The expressions of $\alpha$, $\beta$ in terms of the model parameters can be found, for example, in Ref.~\cite{Coleppa:2013dya, Branco:2011iw}.
It is these neutral scalars $A,H$ that we are studying in this work. 

In some regions of parameter-space, $m_A \approx m_H$, i.e. their masses are within the experimental resolution to distinguish them.
If so, we must add the contributions from both $A$ and $H$ to any given channel; their sum is incoherent due to the different CP quantum-numbers.
For instance, the experimental invariant-mass resolution in the $\tau^{+} \tau^{-}$ channel is about 30\,\% (see for instance Ref.~\cite{Aad:2014vgg}).
Therefore, we consider two cases, one when $m_{A}$ and $m_{H}$ are within 30\,\% and add the contributions from the ``degenerate'' $A$ and $H$, 
and another when they are split by more than $30\,\%$ and treat them separately.
When they are degenerate, for the $\tau^{+} \tau^{-}$ channel for instance, we have $BR(A\to\tau^{+} \tau^{-}) \approx BR(H\to\tau^{+} \tau^{-})$ in the so-called alignment limit (as will be defined precisely later), 
and we can use the constraints obtained in Sec.~\ref{ModInd.SEC} if we interpret $\kappa_{\phi gg}$ shown there as $\sqrt{\kappa_{Agg}^{2} + \kappa_{H gg}^{2}}$ and 
$BR(\phi\rightarrow \tau \tau)$ as $ BR(A\rightarrow \tau^{+} \tau^{-}) + BR(H\rightarrow \tau^{+} \tau^{-})$.
For the non-degenerate case, again one can make use of our results in Sec.~\ref{ModInd.SEC} to obtain constraints either for the $H$ or $A$.

We are interested in the case where the lighter CP-even scalar ($h$) is the observed 125~GeV Higgs boson.
For this, the cos$(\beta-\alpha) \approx 0$ is the most favored region (see Fig.~18 of Ref.~\cite{Dumont:2014kna}).
Only a small range of other values of $(\beta-\alpha)$ are allowed where the sign of the down-type coupling of the Higgs is reversed.
For the 2HDM with exact $Z_2$ symmetry (i.e. $m_{12}^2=0$), $\tan\beta$ has an upper limit of $7$ from perturbativity constraint (see Ref.~\cite{Chen:2013rba}).
We will work with a nonzero $m_{12}^2$ which allows for larger values of $\tan \beta$ (see Ref.~\cite{Das:2015qva}).
We also assume that the ``alignment limit'' ($\beta-\alpha=\pi/2$) holds sufficiently accurately so that the $h$ couplings are SM like
to match with the properties of the observed 125~GeV state at the LHC as discussed in Ref.~\cite{Dumont:2014wha}. 
In this limit, the $H\rightarrow WW$ and $H\rightarrow ZZ$ decays do not give any significant constraints on the parameter space (see for example Ref.~\cite{Chatrchyan:2013yoa}). 

Depending on how the fermions couple to $\Phi_1$ and $\Phi_2$, various types of 2HDM have been defined in the literature, 
some of which we discuss next. 
We start by discussing a 2HDM with only the SM fermions present, and follow it up with 
many examples of different ways of adding vector-like fermions.
Although we do not discuss it here, it is possible in some models for the $y_{Htt}$ coupling to be accidentally suppressed and consequently for the 
$BR(\phi \to t\bar t)$ to become small, as for example in the model discussed in Ref.~\cite{Gopalakrishna:2015dkt}. 

\subsubsection{Type-II 2HDM}
In the Type-II 2HDM the SM Yukawa couplings are replaced by
\beq
  \label{2hdm2}
 \mathcal{L}\supset -y_{d}\bar q_{L} \Phi_{1} d_{R}  - y_{u}\bar q_{L} \tilde{\Phi}_{2} u_{R}  + h.c.  \ ,
\eeq
where $\tilde{\Phi}_i = i \sigma^2 \Phi_i^*$.
The Yukawa couplings of $h,A$ to the SM fermions are given as,
\beq
\label{SMyuk}
\mathcal{L}\supset - \frac{1}{\sqrt{2}}\left( y_u h c_\alpha \bar{u}_L u_R - y_d h s_\alpha \bar{d}_L d_R - y_u c_\beta i A \bar{u}_L u_R - y_d s_\beta i A \bar{d}_L d_R + h.c. \right)\ .
\eeq
The $H$-Yukawa couplings can be obtained from the $h$-Yukawa couplings by the replacements $s_\alpha \to -c_\alpha$ and $c_\alpha \to s_\alpha$.
We find the allowed regions of parameter space from the exclusion-limit on 
$\sigma(gg\to \phi) \times BR (\phi \rightarrow \tau^{+} \tau^{-})$ presented by ATLAS~\cite{atlas_ichep,Aad:2014vgg}.
We focus on the $\tau^{+} \tau^{-}$ channel as currently this is the most constraining one.
We do this first in the 2HDM Type-II (2HDM-II) without the addition of any VLFs.

\begin{figure}[]
\centering
\includegraphics[width=0.32\textwidth]{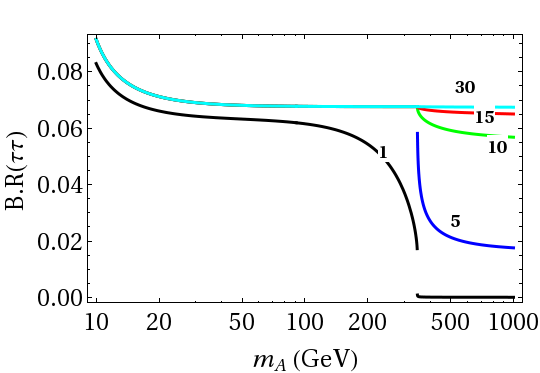}
\includegraphics[width=0.32\textwidth]{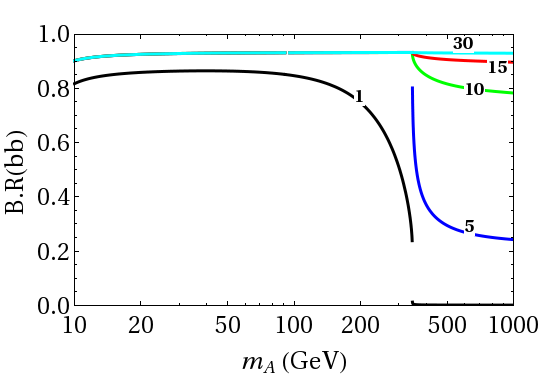}
\includegraphics[width=0.32\textwidth]{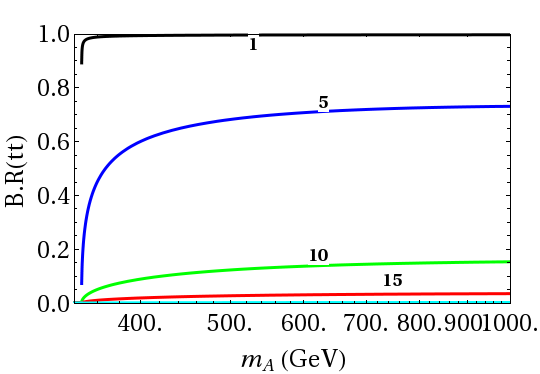}
\caption{$BR(A\to\tau^{+} \tau^{-}, b \bar b)$ (left, middle) for $\tan \beta=1,5,10,15,30$ 
and $BR(A \to t \bar t)$ (right) for $\tan \beta=1,~5,~10,~15$ in 2HDM Type-II model.
The loop-level $BR(A\to VV)$ in the Type-II 2HDM are shown in Fig.~\ref{br_c1} by the dashed-black curves.}
\label{BRA2ff-2HDMII.FIG}
\end{figure}

In Fig.~\ref{BRA2ff-2HDMII.FIG} we show the tree-level decays of $A$ to SM fermions 
BR($A\rightarrow b \bar b, \tau^{+} \tau^{-}, t \bar t$) as a function of $m_A$ for various $\tan\beta$
for the Type-II 2HDM. 
The loop-level $BR(A\to \gamma\gamma, Z\gamma)$ in the Type-II 2HDM are shown in Fig.~\ref{br_c1} by the dashed-black curves,
and our results 
match with that of the Ref.~\cite{Djouadi:2005gj}. 
We see that the BRs into $\gamma \gamma$ and $Z \gamma$ are smaller compared to that of the corresponding loop induced SM Higgs branching ratios even for $\tan \beta = 1$
when the couplings of $A$ to the SM fermions are equal to the Higgs Yukawa couplings.
This is because
the partial width $\Gamma (h\rightarrow \gamma \gamma, \gamma Z$), being dominated by the $W$ loop,
is larger than the partial width $\Gamma(A \rightarrow \gamma \gamma, \gamma Z)$ in which only the fermions contribute (see for example Fig.~2.10 of Ref.~\cite{Djouadi:2005gj}).
For larger $\tan \beta$ the branching ratios are even smaller because of the increased $\Gamma (A\rightarrow b \bar b)$ and $\Gamma(A \rightarrow \tau^{+} \tau^{-} )$ 
(recall that the $Ab \bar b$ and $A \tau^{+} \tau^{-}$ couplings are proportional to $\tan \beta$).
The discontinuity at $m_{A}=2 m_t$ in the BRs in Fig.~\ref{br_c1} for $\tan \beta=1$ is because of the onset of $A \rightarrow t \bar t$ on-shell decay.
For larger $\tan \beta$, the discontinuity is smaller since the $At \bar t$ coupling becomes smaller. 
The $h \rightarrow AA$ decay, possible for $m_{A}< m_{h}/2$, is studied in Ref.~\cite{Dumont:2014kna} and we will not discuss it here.

\begin{figure}[]
\centering
\includegraphics[width=0.32\textwidth]{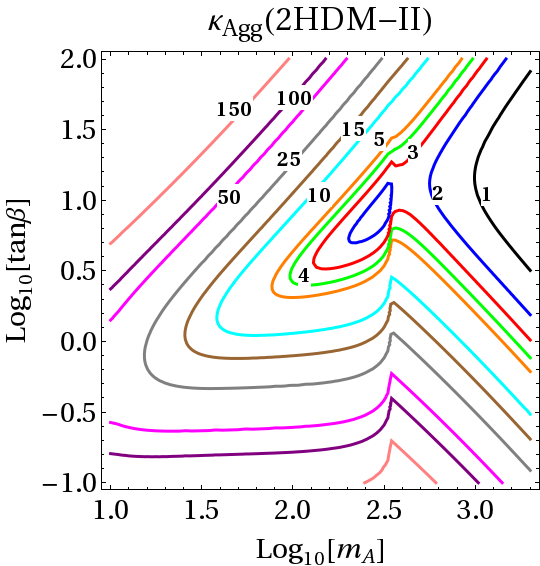}
\includegraphics[width=0.32\textwidth]{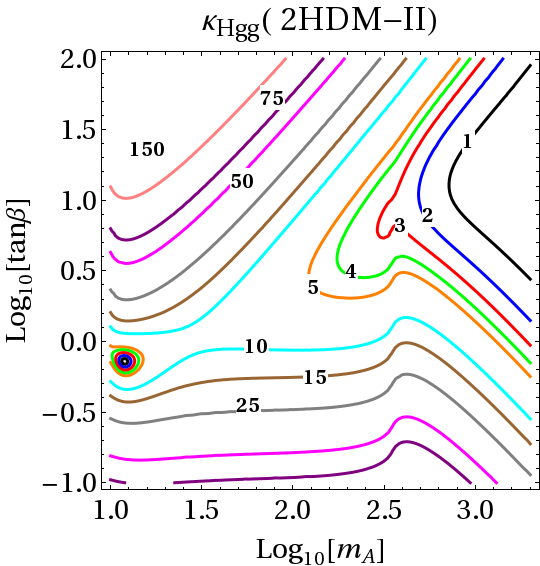}
\caption{Contours of $\kappa_{A gg }$ (left) and $\kappa_{H gg}$ (right) in the Type-II 2HDM.}
\label{kphigg-2HDM-II.FIG}
\end{figure}

\begin{figure}[]
\centering
{}{\includegraphics[width=0.32\textwidth]{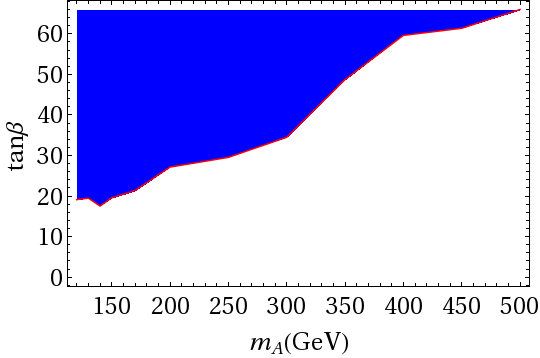}}\hspace{1em}%
{}{\includegraphics[width=0.32\textwidth]{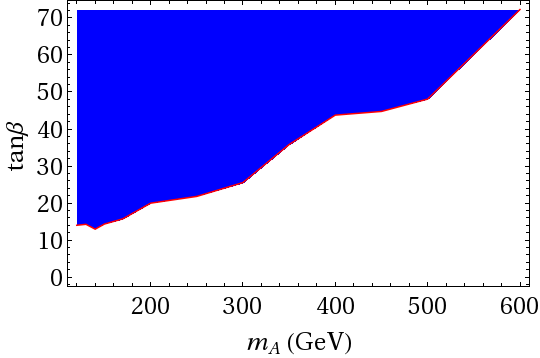}}
\caption{For Type-II 2HDM, regions of the  $m_{A}$-$\tan \beta$ parameter space (blue region) which is excluded at $95 \%$ confidence level 
from $\phi \rightarrow \tau^{+} \tau^{-}$ decay when only $A$ is present (left) 
and when $m_{A}$ and $m_{H}$ are degenerate (right).}
\label{constraints_tanb_2hdm}
\end{figure}

In Fig.~\ref{kphigg-2HDM-II.FIG}, we plot contours of $\kappa_{Agg}$ and $\kappa_{H gg }$ in Type-II 2HDM.
Using this, one can read-off the $\sigma(gg\rightarrow \phi)$ at the 8~and~14~TeV LHC from 
Fig.~\ref{kgg_prod} in Sec.~\ref{ModInd.SEC}.
Using the $\tau^{+} \tau^{-}$ channel constraints shown in Fig.~\ref{kgg_8TeV} of Sec.~\ref{ModInd.SEC} 
we obtain constraints on this model.  
In Fig.~\ref{constraints_tanb_2hdm} we plot the $95\%$ confidence level constraints on the $m_{A}$-$\tan\beta$ plane,
when only $A$ is present (left), and for $m_A = m_H$ when both contribute (right).
Ref.~\cite{Aad:2014vgg} has presented similar constraints in the $m_{A}$--$\tan \beta$ plane, but for the MSSM.

\subsubsection{Type-X 2HDM}
\begin{figure}
\centering
\includegraphics[width=0.32\textwidth]{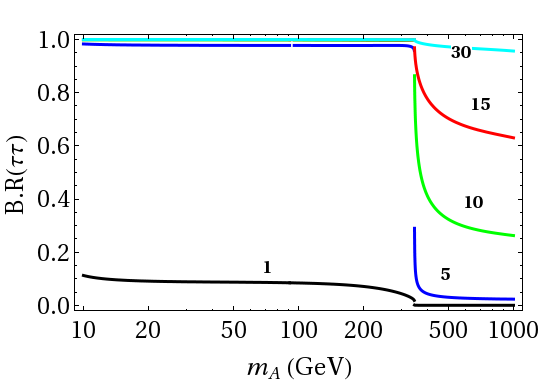}
\includegraphics[width=0.32\textwidth]{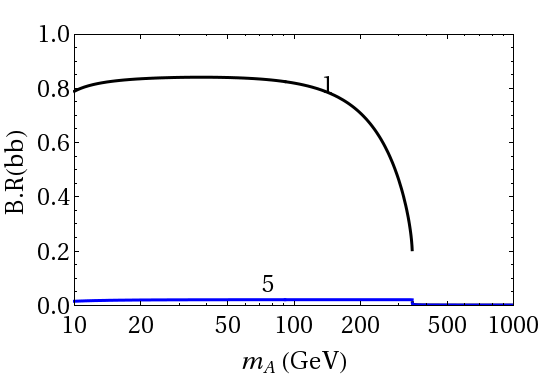}
\includegraphics[width=0.32\textwidth]{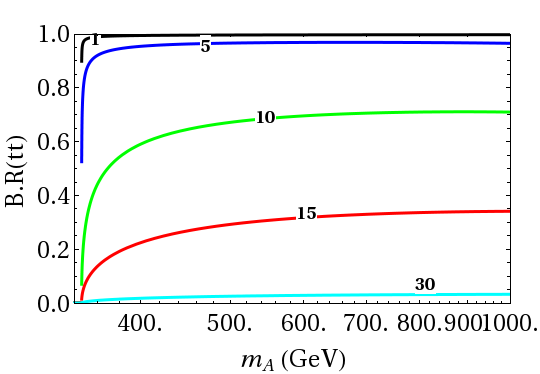}
\caption{For the Type-X 2HDM, 
BR$(A \rightarrow \tau^{+} \tau^{-}, t\bar t)$ (left, right) for $\tan \beta=1,5,10,15,30$, and
BR($A \rightarrow b \bar b$) (middle) for $\tan \beta=1,~5$.
The $BR(A\to \gamma\gamma, Z\gamma)$ for the Type-X 2HDM is shown in Fig.~\ref{br_m3} as the dashed-black curve.
} 
\label{BRA2ff-2HDM-X.FIG}
\end{figure}

In the Type-X 2HDM (2HDM-X) (see Ref.~\cite{Chang:2012ve,Branco:2011iw} for a description of this model) 
all the SM quarks couple to $\Phi_{2}$ and all the leptons couple to $\Phi_{1}$. 
The Lagrangian for the model 2HDM-X is given by
\begin{align}
\label{L2HDMX}
 \mathcal{L}\supset &-(y_{d}\bar 
 q_{L} \Phi_{2} d_{R}  + y_{u}\bar q_{L} \tilde{\Phi}_{2} u_{R} + y_{e}\bar 
 l_{L} \Phi_{1} e_{R} + h.c) + \left| D_{\mu}\Phi_{1} \right|^{2} + \left|D_{\mu}\Phi_{2}\right|^{2}
                    -V(\Phi).
\end{align}
As a result, $A$ coupling to the quarks and leptons are  proportional to $\cot\beta$ and $\tan \beta$ respectively. 
In the Type-X model, since all SM quarks couple very weakly to $A$ for large $\tan \beta$, $\sigma(gg\rightarrow A)$ becomes very small for large $\tan\beta$.
As a consequence there are no constraints from $\sigma(pp \rightarrow A)\times BR (A \rightarrow \tau^{+} \tau^{-})$.
The SM quark contribution to $\kappa_{AVV}$ for 2HDM-X can obtained from that of 2HDM-II (see Ref.~\cite{Djouadi:2005gj}) 
by replacing $\tan \beta$ with $\cot \beta$ in the $Ab \bar b$ coupling.
In Fig.~\ref{BRA2ff-2HDM-X.FIG} we show the tree-level BR($A \rightarrow \tau^{+} \tau^{-}, b \bar b, t \bar t$).
The $BR(A\to VV)$ for the Type-X 2HDM is shown in Fig.~\ref{br_m3} as the dashed-black curve. 

\begin{figure}[]
\centering
\includegraphics[width=0.32\textwidth]{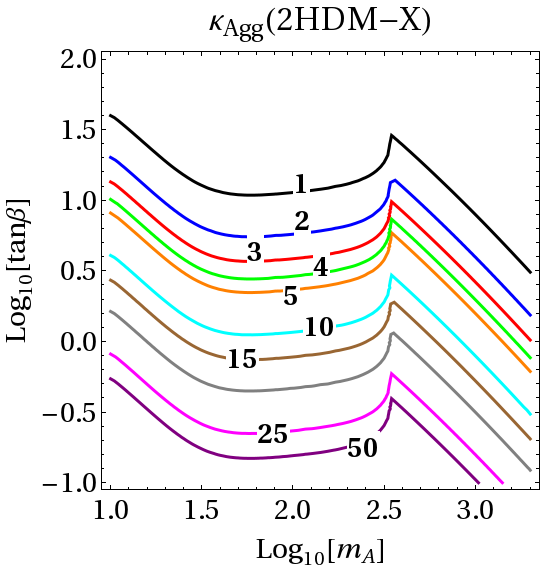}
\includegraphics[width=0.32\textwidth]{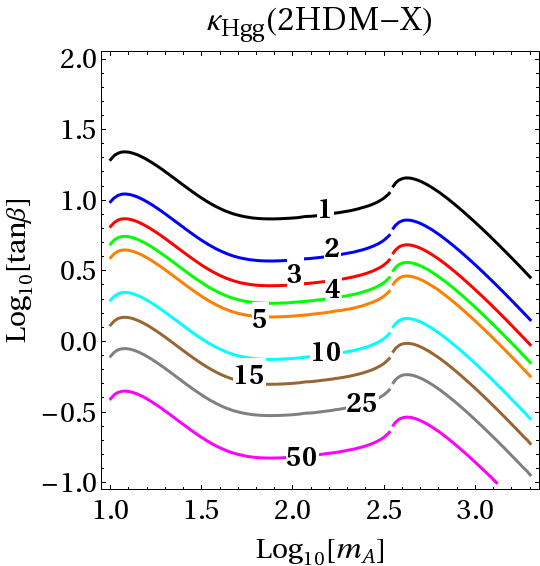}
\caption{For the Type-X 2HDM, contours of $\kappa_{A gg }$ (left) and $\kappa_{Hgg}$ (right).}
\label{kphigg-2HDM-X.FIG}
\end{figure}

In Fig.~\ref{kphigg-2HDM-X.FIG} we plot contours of $\kappa_{Agg}$ and $\kappa_{Hgg}$.
From this one can read off $\sigma(gg \rightarrow \phi)$ for 8~TeV and 14~TeV LHC from Fig.~\ref{kgg_prod} 
in Sec.~\ref{ModInd.SEC}. 
The results for $\kappa_{\phi gg}$ in 2HDM-X are also applicable for Type-I 2HDM
as the SM quarks couple to $H,A$ in an identical fashion as in Type-X 2HDM.

Next, we add various combinations of $SU(2)$ singlet and doublet VLFs to the Type-II 2HDM first, and to Type-X 2HDM 
following that.
Our goal is to study how VLFs affect the LHC production rate and decay BRs of the $\phi$.
There are eight different ways in which the $\Phi_{1}$ and the $\Phi_{2}$ can couple to the VLFs consistent with the symmetries of the 2HDM-II namely 
$\Phi_{1}\rightarrow -\Phi_{1}$ and $d_{R}\rightarrow -d_{R}$ (with all other fields unchanged).
Among these eight models we will discuss only three representative ones that also capture the effects in the others.

\subsubsection{Type-II 2HDM with VLQ-SMQ Yukawa couplings}
\label{VLF.mix}
Many models that address the hierarchy problem, such as for example composite-Higgs and little-Higgs models,
have as an important ingredient off-diagonal couplings between a VLF and 3rd-generation SM fermions. 
We discuss this possibility in a model-independent way by introducing, 
one at a time, SU(2)-singlet VLFs with EM charge $2/3$ and $-1/3$.
As an example, we show how the results obtained here apply to a little-Higgs model.
%

\medskip
\noindent \underline{\bf $MVU$ model}:
In what we call the $MVU$  model, we introduce an SU(2)-singlet VLF pair $(\psi, \psi^c)$, 
denoted by the 4-spinor $\psi$, with EM charge 2/3, 
and add to the 2HDM Type-II Lagrangian the following terms 
\beq
{\cal L} \supset M_\psi \, \bar\psi \psi - \left( y_1 \, \bar q_L \tilde{\Phi}_1 \psi_R  + h.c. \right) 
\ . 
\label{LtpSing.EQ}
\eeq
After EWSB the mass terms for the EM-charge 2/3 fermions can be written as
\begin{equation}
  \mathcal{L}^{mass} =- \frac{1}{\sqrt{2}}  \left(y_u v_2 \bar t_L t_R + y_1 v_1 \bar t_L \psi_R + h.c.\right)  + M_{\psi} \bar{\psi} \psi .
\end{equation}
We define the mass eigenstates $t^0_{L,R}$ and $t_{2L,R}$, for the EM-charge 2/3 quarks as 
\begin{align}
\label{dmixing}
 &t_{L,R}= \cos \theta^U_{L,R} t^0_{L,R} - \sin \theta^U_{L,R} t_{2L,R}, \nonumber \\
 &\psi_{L,R}= \sin \theta^U_{L,R} t^0_{L,R} + \cos \theta^U_{L,R} t_{2L,R}.
\end{align}
The mixing angles and the mass eigenvalues can be found in App.~\ref{tpbpMangle.App}.
For notational brevity we call $t^0$ simply as $t$, which we will identify with the SM top-quark. 
Constraints on the mixing from EWPT and a vector-like top decaying to $W b,~Z t,~ H t$ are studied in 
Refs.~\cite{VLFDir.REF,Ellis:2014dza,Cacciapaglia:2011fx,Dawson:2012di}.
Constraints from flavor observables are studied in Ref.~\cite{Cacciapaglia:2011fx}.

\begin{figure}[h]
\centering
{}{\includegraphics[width=0.32\textwidth]{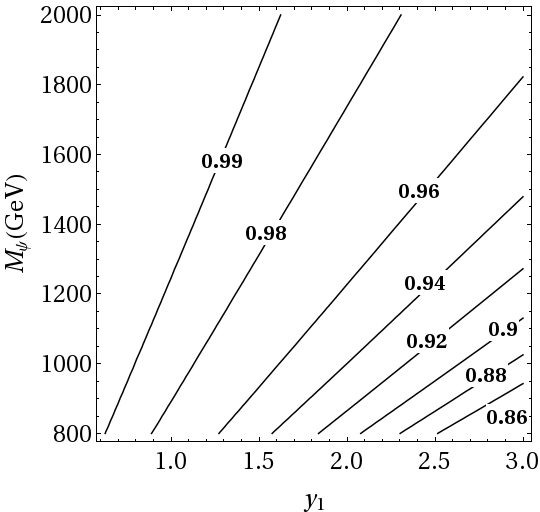}}
{}{\includegraphics[width=0.32\textwidth]{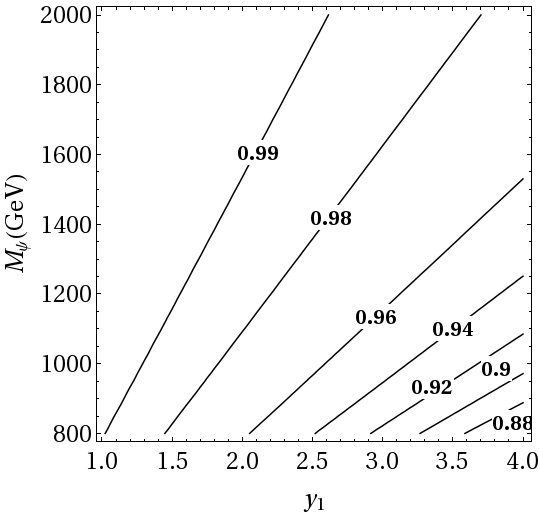}}
\caption{For the $MVU$ model, contours of $\kappa_{htt}$ for $\tan \beta = 1$ (left) and $\tan \beta = 5 $ (right) 
with $y_u$ chosen such that $m_t = 163~$GeV.}
 \label{mtsmytsm-tp}
\end{figure}

\begin{figure}[h]
\centering
{}{\includegraphics[width=0.3\textwidth]{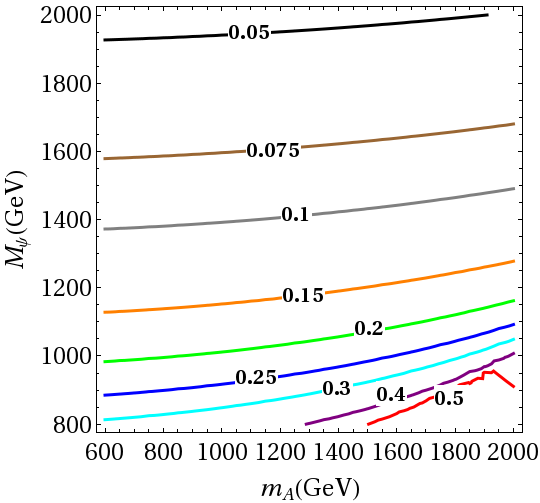}}
{}{\includegraphics[width=0.3\textwidth]{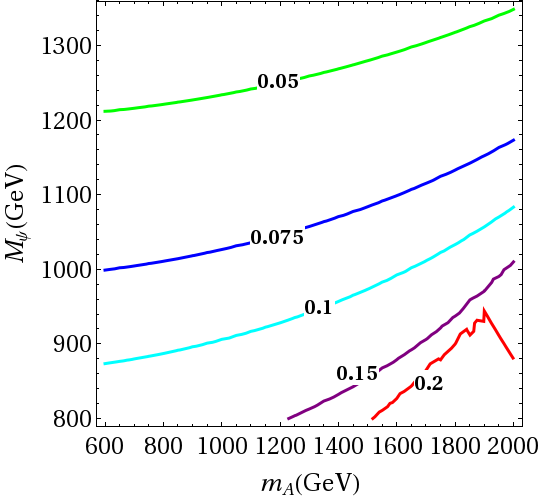}}
{}{\includegraphics[width=0.3\textwidth]{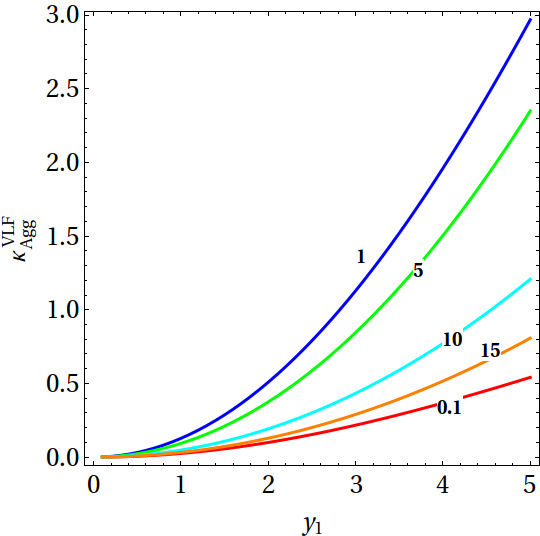}}
\caption{Contours of $k_{Agg}^{\text{VLF}}/y_1^2$ for $\{\tan \beta,~y_u\} = \{1,~1.4 \}$ (left) and $\{ 5,~1 \}$ (middle) 
for the $MVU$ model.
$k_{Agg}^{\text{VLF}}$ as a function of $y_1$, for $m_A=1000~$GeV, $M_\psi= 1250~$GeV and $\tan \beta = 0.1,1,5,10,15$ 
is plotted on the right.}
 \label{kAggvlf-tp}
\end{figure}

The $A$ couplings to the EM-charge 2/3 fermions in terms of the mass eigenstates are given by
\begin{equation}
\label{Add.Mixing}
 \mathcal{L} = \frac{i}{\sqrt{2}}A \left( y_{Att} \bar{t}_L t_R + y_{A t_2 t_2} \bar{t}_{2L} t_{2R} + y_{A t_2 t} \bar{t}_{2L} t_R + y_{A t t_2} \bar{t}_L t_{2R} \right) + h.c. \ ,
\end{equation}
where $y_{A t t} \supset (y_u c^U_L c^U_R \cos \beta - y_1 c^U_L s^U_R \sin \beta) $,
$y_{A t_2 t_2}=(y_u s^U_L s^U_R \cos \beta + y_1 s^U_L c^U_R \sin \beta)$,
$y_{A t_2 t}=-(y_u s^U_L c^U_R \cos \beta - y_1 c^U_L s^U_R \sin \beta)$, and
$y_{A t t_2}=-(y_u c^U_L s^U_R \cos \beta + y_1 c^U_L c^U_R \sin \beta)$.
The $h$ couplings to the EM-charge 2/3 fermions are given by,
\begin{equation}
\label{hdd.Mixing}
 \mathcal{L} \supset \frac{1}{\sqrt{2}} h \left( y_{htt} \bar{t}_L t_R + y_{h t_2 t_2} \bar{t}_{2L} t_{2R} + y_{h t_2 t} \bar{t}_{2L} t_R + y_{h t t_2} \bar{t}_L t_{2R} \right) + h.c. \ ,
\end{equation}
where $y_{h t t}=(- y_u c^U_L c^U_R \cos \alpha + y_1 c^U_L s^U_R \sin \alpha) $,
$y_{h t_2 t_2}=(-y_u s^U_L s^U_R \cos \alpha - y_1 s^U_L c^U_R \sin \alpha)$,
$y_{h t_2 t}=(y_u s^U_L c^U_R \cos \alpha - y_1 c^U_L s^U_R \sin \alpha)$, and 
$y_{h t t_2}=(y_u c^U_L s^U_R \cos \alpha +  y_1 c^U_L c^U_R \sin \alpha)$.
We fix $m_t^{\bar{MS}} = 163~$GeV~\cite{Alekhin:2012py} by choosing $y_u$ appropriately, 
and show in Fig.~\ref{mtsmytsm-tp} the contours of $\kappa_{htt}\equiv y_{htt}/y^{SM}_{htt}$ in the $y_1$--$M_\psi$ plane.
In the region to the left of the $0.99$ contours, $\kappa_{htt}$ approaches $1$.
The experimental constraint on $\kappa_{htt}$ is $0.63 < \kappa_{htt} < 1.2$~\cite{hCouplComb.BIB}.
In Fig.~\ref{kAggvlf-tp} we show contours of $\kappa_{Agg}^{\text{VLF}}/y_1^2$ in the $m_A$--$M_\psi$ plane 
for $\{\tan \beta, y_u\} = \{1, 1.4\}$ and $\{ 5, 1\}$,
and also show $\kappa_{Agg}^{\text{VLF}}$ as a function of $y_1$ 
for $m_A = 1000~$GeV, $M_\psi=1250~$GeV and $\tan \beta = 0.1~,1,~5,~10,~15$.
For large $\tan \beta$, the mixing angles become small, which makes $\kappa_{Agg}^{\text{VLF}}$ small. 
For Fig.~\ref{kAggvlf-tp}, we fix $y_u = 1.4$ so that $m_t$ is close to its experimental value, 
and once a specific choice of $y_1$ is made, $m_t$ can be fixed exactly by choosing $y_u$ slightly differently;
the resulting change in $\kappa_{Agg}^{VLF}$ due to such differences in $y_u$ is insignificant. 

\begin{figure}[h]
\centering
{}{\includegraphics[width=0.3\textwidth]{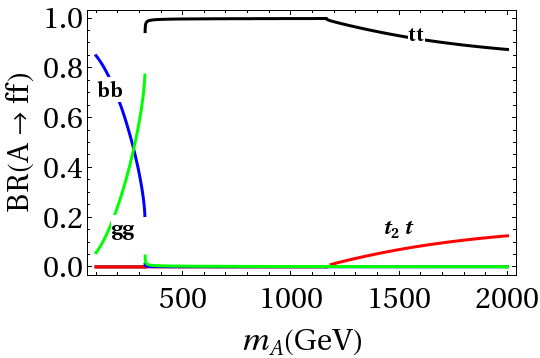}}
{}{\includegraphics[width=0.3\textwidth]{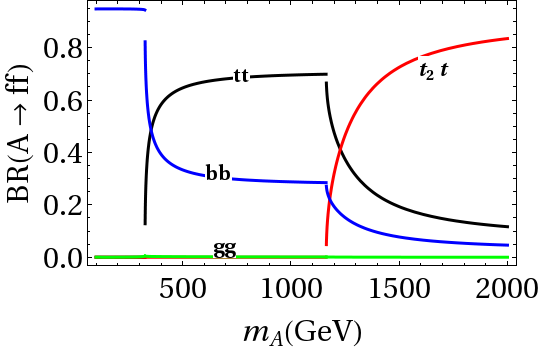}}
\caption{Contours of BR$(A \to tt)$ (black), BR$(A \to bb)$ (blue), BR$(A \to t_2 t)$ (red), BR$(A \to gg)$ (green) with $M_\psi = 1000~$GeV, $y_1=1$ for $ \tan \beta = 1$ (left) and $ 5$ (right), for the $MVU$ model, with $y_u$ chosen such that $m_t=163~$GeV.}
 \label{brAff-mA-mvu}
\end{figure}

The fermionic decay BR for $m_A < (M_{t_2}+m_t)$ will be largely unchanged from the Type-II 2HDM plots 
shown in Fig.~\ref{BRA2ff-2HDMII.FIG}.  
However, if $m_A > (M_{t_2} + m_t)$ the $A\to t_2 t$ decay becomes kinematically allowed.
In Fig.~\ref{brAff-mA-mvu} we plot BR$(A \to t t)$, BR$(A \to bb)$, BR$(A \to gg)$ and BR$(A \to t_2 t)$, for $M_\psi = 1~$TeV, $y_1=1$ and $\tan\beta=\{1,5\}$ 
with $y_u$ fixed such that $m_t$ is at the physical value.
BR$(A \to \gamma \gamma,~Z \gamma)$ do not change by much from the 2HDM-II case.

As an example, we apply these results to a concrete model that stabilizes the electroweak scale, 
has a 2HDM structure, and has vector-like fermions,
namely the SU(6)/Sp(6) little-Higgs model by Low, Skiba and Smith~(LSS, Ref.~\cite{Low:2002ws}), 
which we analyze in detail in Ref.~\cite{Gopalakrishna:2015dkt}.
Among the various sample points that are listed in App.~B in Ref.~\cite{Gopalakrishna:2015dkt} that satisfy all constraints including precision electroweak, we consider here the sample-points 1~and~2.
For the sample-point-1,
the two lightest VLFs are the $t_2$ with a mass of $1218~$GeV, the $b_2$ with a mass of $1315$~GeV,
and we have $\tan\beta = 1.36$, $m_A=1671~$GeV, $y_1 = 1.7$, 
$y_u = 1.2$ and $m_t \approx 164~$GeV.\footnote{From Ref.~\cite{Gopalakrishna:2015dkt} we have the following for the LSS model:
there we had $\tan \beta = v_1/v_2$ while in this paper we have $\tan \beta = v_2/v_1$, and therefore
$\tan \beta$ here is related to that of Ref.~\cite{Gopalakrishna:2015dkt} via $\tan \beta = (1/\tan \beta^{LSS})$;
$y_1$ is given by $y_1 = y_1^{LSS} c_{23}$, and 
for point-1, since $y_1 \gg y_4$, to a very good approximation $s_{14} \approx 1$ and $c_{14} \approx 0$;
also $m_t \approx c_{23} y_2 v_2/\sqrt{2}$ in the limit where $t_3$ is decoupled away, i.e. $y_u = y_2 c_{23}$, 
and $c_{23} \approx 0.9$. 
The $b_2$ is an SU(2)-singlet since it does not mix with the other charge -1/3 states.}  
Keeping only the lighter $t_2$ since the $t_3$ is somewhat heavier, a good approximation is obtained by considering the 
addition of only a singlet EM charge $+2/3$ state $\psi$ as introduced in Eq.~(\ref{LtpSing.EQ}). 
Ignoring the smaller $b_2$ contribution, the $\kappa_{A gg}^{VL}$ due to the $t_2$ can be read off from the 
$\tan \beta =1$ curve of the rightmost panel of Fig.~\ref{kAggvlf-tp} to be approximately $0.4$.
This is about 10\% of the SM-fermion contribution. 

\begin{figure}[h]
\centering
{}{\includegraphics[width=0.3\textwidth]{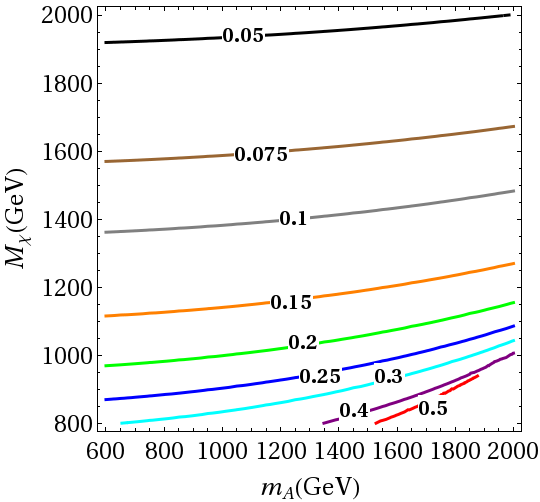}}
{}{\includegraphics[width=0.3\textwidth]{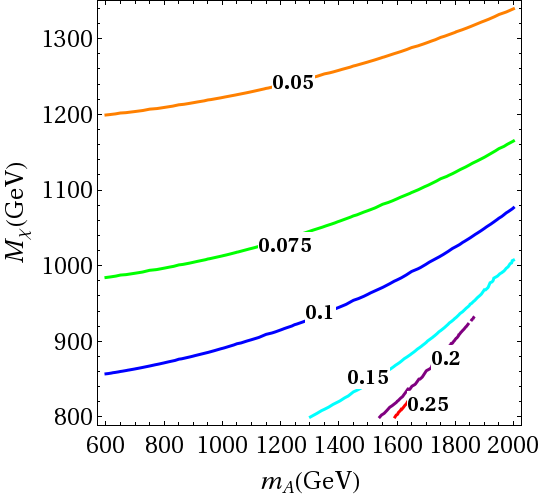}}
{}{\includegraphics[width=0.28\textwidth]{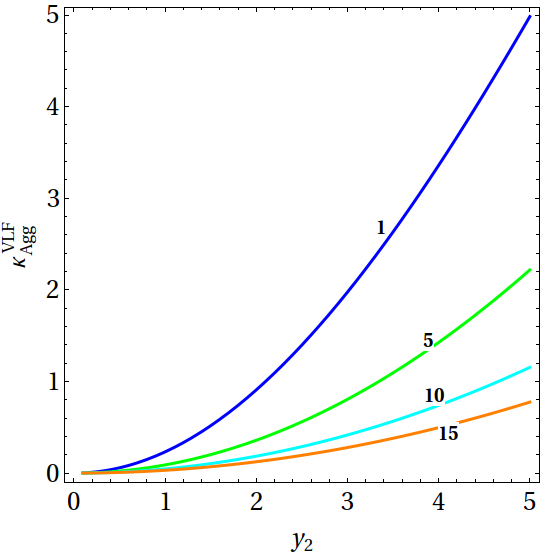}}
\caption{Contours of $k_{Agg}^{\text{VLF}}/y_2^2$ for $\{\tan \beta,~y_d\} = \{1,~0.03 \}$ (left) and $\{ 5,~0.12 \}$ (middle) 
for the $MVD$ model.
$k_{Agg}^{\text{VLF}}$ with $y_2$ for $m_A=1500~$GeV, $M_\chi= 1000~$GeV and $\tan \beta = 1,~5,~10,~15$ is plotted on the right.}
 \label{kAggvlf-bp}
\end{figure}

\medskip
\noindent \underline{\bf MVD model}:
In the $MVD$ model, we introduce an SU(2)-singlet VLF pair $(\chi, \chi^c)$, denoted by the 4-spinor $\chi$, 
with EM charge -1/3, and add to the 2HDM Type-II Lagrangian the following terms 
\beq
{\cal L}_{A} = M_\chi \, \bar\chi \chi - \left( y_2\, \bar q_L \Phi_1 \chi_R + h.c. \right)  \ .
\eeq
The mass eigenstates, $b^0_{L,R}$ and $b_{2L,R}$ for the EM-charge $-1/3$ fermions are defined in the same way as in 
Eq.~(\ref{dmixing}) with the mixing angles, $\theta^D_{L,R}$.
The mixing angles and the mass eigenvalues can be found in App.~\ref{tpbpMangle.App}.
The $A$ couplings to the EM-charge $-1/3$ fermions are obtained in a similar way as in Eq.~(\ref{Add.Mixing}),
with the replacements $y_u \cos \beta  \to y_d \sin \beta$, $y_1 \to  y_2$.
Similarly, the $h$ couplings to the EM-charge $-1/3$ fermions are obtained from Eq.~(\ref{hdd.Mixing}),
with the replacements $y_u \cos \alpha \to - y_d \sin \alpha$ and  $y_1 \to y_2$.
As in the case of charge 2/3 fermions, we choose $y_d$ 
such that $m_b^{\bar{MS}} = 4.2$~GeV~\cite{Agashe:2014kda}; $y_{hbb}$ stays close it's SM value.
In Fig.~\ref{kAggvlf-bp} we plot contours of $\kappa_{Agg}^{\text{VLF}}/y_2^2$ in the $m_A - M_\chi$ plane for 
$\{\tan \beta,~y_d \} = \{1,~0.03 \}~{\rm and}~\{ 5,~0.12 \}$, and 
$\kappa_{Agg}^{\text{VLF}}$ as a function of $y_2$ for $m_A = 1500~$GeV, $M_\chi=1000~$GeV for $\tan \beta = 1,~5,~10,~15$.

As an example we consider again the LSS model, but now the sample-point-2 in App.~B of Ref.~\cite{Gopalakrishna:2015dkt}, 
with the lightest VLF being the $b_2$ with a mass of $947.5~$GeV. 
The $b_2$ is an SU(2)-singlet state and does not mix with the other charge -1/3 states. 
For this point, $m_A=1671~$GeV, $\tan\beta = 1.36$, $y_2 = 1.422, c_{23}=1.15$. 
$\kappa_{A gg}$ can be read off from the $\tan \beta =1$ curve of the rightmost panel of Fig.~\ref{kAggvlf-bp} 
to be approximately $0.3$.

\medskip
\noindent \underline{\bf $MVQ$ model}:
For the $MVQ$ model, we add an SU(2) doublet VLF pair $(Q',{Q'}^c)$ denoted by the 4-spinor $Q'$, 
and add to the Type-II 2HDM Lagrangian the terms
\beq
{\cal L} \supset M_{QQ}\, \bar{Q}' Q' + \left( M_{qQ}\, \bar{q}_L Q'_R - \tilde{y}_1 \bar{Q}'_L \tilde{\Phi}_2 t_R -\tilde{y}_2 \bar{Q}'_L \Phi_1 b_R + h.c. \right) \ .
\eeq
In the following we show only the top-sector since this is usually the dominant piece in BSM models, 
and we therefore suppress the bottom-sector. 
At the outset, we diagonalize the VLF masses by redefining the $Q$ and $Q'$ fields 
by an orthogonal rotation to get an equivalent Lagrangian given by
\beq
{\cal L} \supset M^{\rm\it eff}_{QQ}\, \bar{Q}' Q' + \left(-y^{\rm \it eff}_u \bar{q}_L \tilde{\Phi}_2 t_R 
                                     - \tilde{y}^{\rm \it eff}_1 \bar{Q}'_L \tilde{\Phi}_2 t_R  + h.c. \right) \ ,
\eeq 
where we show the 2HDM top Yukawa coupling also since its effective coupling is now changed, with 
$M^{\rm\it eff}_{QQ} \equiv \sqrt{(M_{QQ}^2 + M_{qQ}^2)}$, 
$y_u^{\rm\it eff} \equiv (y_u M_{QQ} - \tilde{y}_1 M_{qQ})/M^{\rm\it eff}_{QQ}$, 
$\tilde y_1^{\rm\it eff} \equiv (y_u M_{qQ} + \tilde{y}_1 M_{QQ})/M^{\rm\it eff}_{QQ}$, 
which imply
$y_u^{\rm\it eff} = (y_u-\tilde{y}_1 M_{qQ}/M_{QQ})/\sqrt{1+(M_{qQ}/M_{QQ})^2}$ and
$\tilde y_1^{\rm\it eff} = y_u^{\rm\it eff} M_{qQ}/M_{QQ} + \tilde{y}_1 \sqrt{1+(M_{qQ}/M_{QQ})^2}$.
 
The $\kappa_{\phi gg}$ due to the $t',b'$ in the $MVQ$ model are qualitatively similar 
to the $MVU$ case presented earlier.  
As an example, let us consider again the LSS model sample-point-1 in App.~B of Ref.~\cite{Gopalakrishna:2015dkt}, 
for which we have 
$\tilde {y}_1 = 0$,
$y_u^{\rm\it eff} \approx 1.3$ and 
$\tilde y_1^{\rm\it eff} \approx 0.5$, which gives $\kappa_{Agg} \approx 0.03$. 
The doublet-VLQ contribution in this case is thus very small compared to the SMQ contribution.

\subsubsection{Type-II 2HDM with VLQ-VLQ Yukawa couplings}
Here, we add SU(2) doublet and singlet VLFs with SM-like hypercharge assignments, and write Yukawa couplings between them both involving the $\Phi_{1,2}$.
Although there could be Yukawa couplings between a VLF and an SMF also present, we do not write them here for simplicity,
and their effects are investigated separately in Sec.~\ref{VLF.mix}. 

\medskip
\noindent \underline{\bf $MVQD_{11}$ model}: 
To the Type-II 2HDM we introduce one doublet VLQ, $\psi=(\psi_{1}, \psi_{2})$ with hypercharge $Y_{\psi}$
and one singlet VLQ ($\chi$) with hypercharge $(Y_{\psi}-1/2)$ so that VLF couplings with $\Phi_{1}$ are allowed.
The additional Lagrangian terms to the 2HDM-II are
\begin{align}\label{case2a}
 \mathcal{L}\supset \bar{\psi} i\slashed D \psi + \bar{\chi} i\slashed D  \chi -  ( {y_{1}}\bar \psi_{L} \Phi_{1} \chi_{R}  + \tilde{y}_{1}\bar \psi_{R} \Phi_{1} \chi_{L} +\text{h.c})
                    -M_{\psi} \bar \psi \psi -M_{\chi}\bar \chi \chi .
\end{align}
We can also write 
the terms $\bar \psi_{L} \Phi_{2}\chi_{R}$ and $\bar \psi_{R} \Phi_{2} \chi_{L}$,
which we do not add here but will consider them subsequently as another model.
These terms are forbidden if $\chi \to -\chi$ under the $Z_{2}$ symmetry of 2HDM-II.
The terms involving $h$, $A$ and VLFs after EWSB are  
\begin{align}{\label{model2aA}}
\mathcal{L} \supset & - M_{\psi} \bar \psi \psi - M_{\chi} \bar \chi \chi + \frac{1}{\sqrt{2}} A \sin \beta (i y_{1} \bar \psi_{2L} \chi_{R} + i\tilde{y}_{1} \bar \psi_{2 R} \chi_{L} + h.c.)  
             - \frac{v}{\sqrt{2}} \cos \beta (y _{1}\bar \psi_{2L} \chi_{R} + \tilde{y}_{1}\bar \psi_{2R} \chi_{L} + h.c.)   \nonumber \\
             & +\frac{1}{\sqrt{2}} h \sin \alpha(y_{1} \bar \psi_{2L} \chi_{R} + \tilde{y}_{1} \bar \psi_{2R} \chi_{L} + h.c.) - \frac{1}{\sqrt{2}}H \cos \alpha(y_{1} \bar \psi_{2L} \chi_{R} + \tilde{y}_{1} \bar \psi_{2R} \chi_{L} + h.c.)  \ .
\end{align}
Gauge interactions of the VLFs are present and not shown explicitly. 
$\psi_{2}$ and $\chi$ mix after EWSB, while $\psi_{1}$ is itself a mass eigenstate. We define the mass eigenstates $\zeta_{1}$ and $\zeta_{2}$ as
\begin{align}
 & \psi_{2L,R}= \zeta_{1L,R} \cos \theta_{L,R} - \zeta_{2L,R} \sin \theta_{L,R}, \\
 & \chi_{L,R}= \zeta_{1L,R} \sin \theta_{L,R} + \zeta_{2L,R} \cos \theta_{L,R}, 
\end{align}
where the mixing angles $\theta_{L}$ and $\theta_{R}$ are defined in App.~\ref{effcoup_c1_append}. 
In terms of these mass eigenstates, the Lagrangian in Eq.~(\ref{model2aA}) can be written as
\begin{align}\label{effcoupc1}
 \mathcal{L} \supset & - y_{ij}^{A} (i A \bar \zeta_{iL} \zeta_{jR} + \text{h.c}) - M_{i}\bar \zeta_{i} \zeta_{i} -M_{\psi} \bar \psi_{1} \psi_{1} + \kappa_{ij}  Z_{\mu} \bar \zeta_{i}  \gamma_{\mu} \zeta_{j}                        
  + eQ_i A_{\mu} \bar \zeta_{i}  \gamma_{\mu} \zeta_{i}  \nonumber \\
                     &-y^{h}_{ij}(h \bar \zeta_{iL} \zeta_{jR} + h.c) -y^{H}_{ij}(h \bar \zeta_{iL} \zeta_{jR} + h.c) \ ,
\end{align}
where $i,j=1,2$ and $y^\phi_{ij}$'s are given in App.~\ref{effcoup_c1_append}.
We take the $y_{1}$ and $\tilde{y}_{1}$ to be real, enforcing CP invariance in the BSM sector.
The relative sign between $y_{1}$ and $\tilde{y}_{1}$ in Eq.~(\ref{case2a}) is physical for the following reason.
If we want to get rid of this relative sign we need to make the transformations $\chi_{L}\rightarrow -\chi_{L}$ and $\chi_{R} \rightarrow  \chi_{R}$, or $\chi_{L}\rightarrow \chi_{L}$ and $\chi_{R} \rightarrow  - \chi_{R}$.
In either case, the $M_{\chi}$ changes its sign and is therefore a physical effect.
For chiral fermions, the sign of the mass term is not physical since one can rotate it away by the above transformations.

Instead of the $\chi$ (with hypercharge $(Y_{\psi}-1/2)$), if we consider a VLF (say $\xi$) of hypercharge $(Y_{\psi}+1/2)$,
we get a different model where the $\xi$ couples to the $\tilde{\Phi}_{1}$ instead of the $\Phi_{1}$.
This model will have similar phenomenology as $MVQD_{11}$~model, which we discuss later.  
 
The effective couplings for this model are given in App.~\ref{kAvvGen.App}.
When $y_{1} = \tilde{y}_{1}$, in addition to $CP$ invariance, the Lagrangian in Eq.~(\ref{model2aA}) is also invariant under $P$ and $C$ individually, with $A$ transforming as
$ A \xrightarrow{P} A,\ A \xrightarrow{C} -A$.
This implies that the VLF contribution to $\kappa_{A VV}$ is zero since $A V_{\mu \nu} \tilde{V}^{\mu \nu}$ is not $P$ invariant (although it is $CP$ invariant). 
Also, the VLF contributions are maximum for $M_{\psi}=M_{\chi}$ when the mixing between the VLFs ($\psi_{2}$ and $\chi$) is maximum. We will take $M_{\psi}$ and $M_{\chi}$ to be equal from now on. 

\begin{figure}[]
\centering
{}{\includegraphics[width=0.32\textwidth]{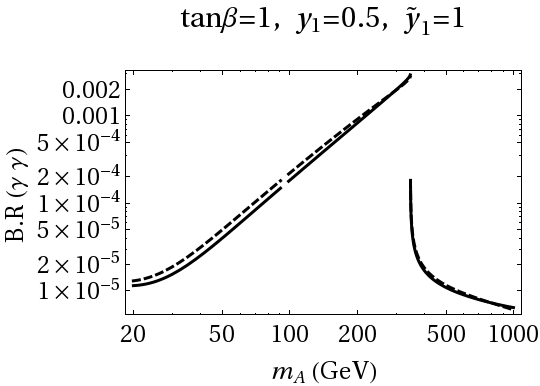}}
{}{\includegraphics[width=0.32\textwidth]{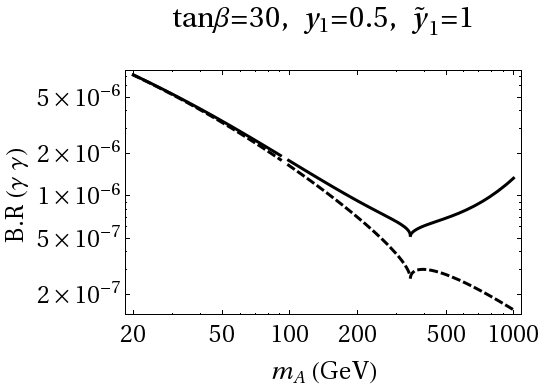}}\\
{}{\includegraphics[width=0.32\textwidth]{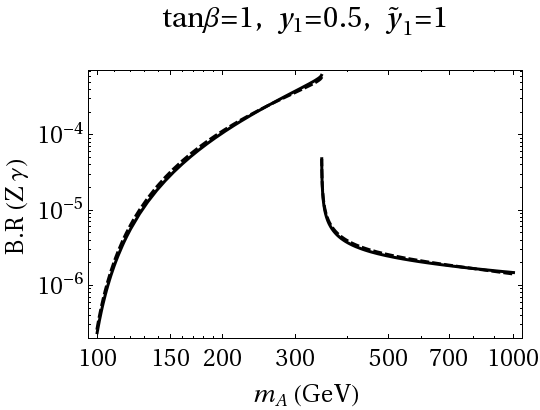}}
{}{\includegraphics[width=0.32\textwidth]{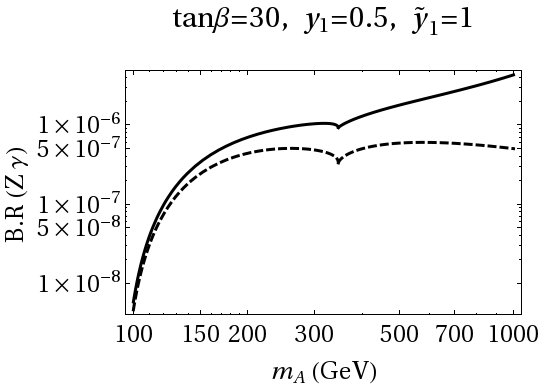}}
\caption{BR($A \rightarrow \gamma \gamma$) (top-panel) and BR($A \rightarrow Z \gamma$) (bottom-panel) 
with $M_{\psi}= M_{\chi}=1000$ (GeV), $\tan \beta = 1$ (left) and $30$ (right) in $MVQD_{11}$~model (solid-black) and in 2HDM Type-II (dashed-black).
$BR(A\to f\bar f)$ in Type-II 2HDM are as shown in Fig.~\ref{BRA2ff-2HDMII.FIG}.}
\label{br_c1}
\end{figure}

In Fig.~\ref{br_c1}, we plot BR($A\rightarrow V V$), for $Y_{\psi}=1/6$ as an example, 
which is the SM quark-doublet hypercharge assignment.
The tree-level decays to SM fermions BR($A\rightarrow b \bar b, \tau^{+} \tau^{-}, t \bar t$) 
are unchanged from what is shown in Fig.~\ref{BRA2ff-2HDMII.FIG} for the Type-II 2HDM. 
We see that for small values of $\tan \beta$ the VLF contribution to BR$(A\rightarrow V V)$ is small compared to the 2HDM-II.
This is because $y_{ij}$'s are proportional to $\sin \beta$. 
For large $\tan \beta$ and for large $m_{A}$, the VLF contributions to the BR($A \rightarrow \gamma \gamma$) become significant. 

\begin{figure}[]
\centering
\includegraphics[width=0.32\textwidth]{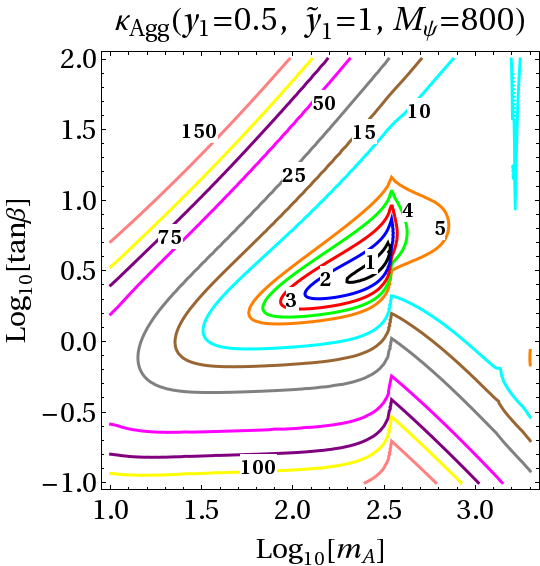}
\includegraphics[width=0.32\textwidth]{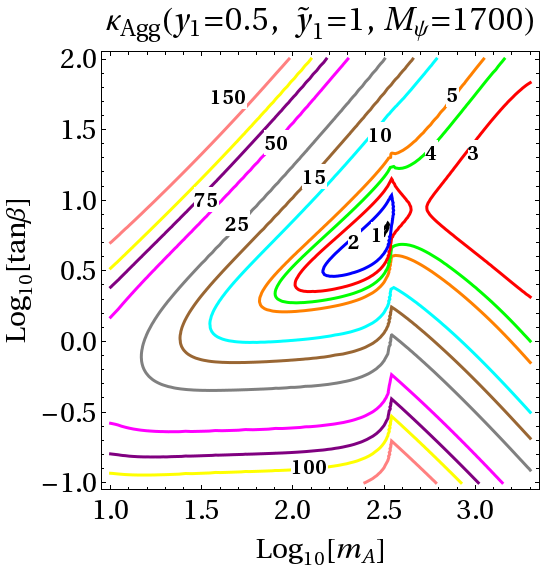}
\caption{Contours of $\kappa_{A gg }$ for $M_{\psi} = M_{\chi}=800~$GeV (left), $1700~$GeV (middle), $y_{1} = 0.5$, $\tilde{y}_{1} =1$ for $MVQD_{11}$~model.}
\label{effcouplingd}
\end{figure}

\begin{figure}[]
\centering
\includegraphics[width=0.32\textwidth]{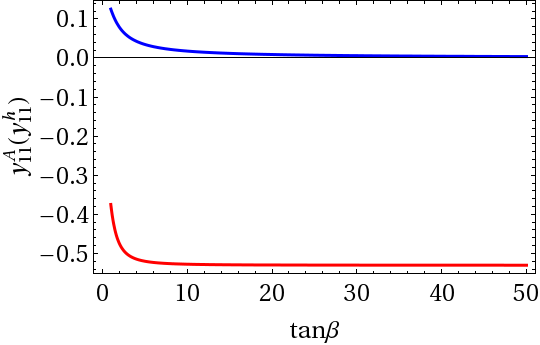}
\includegraphics[width=0.32\textwidth]{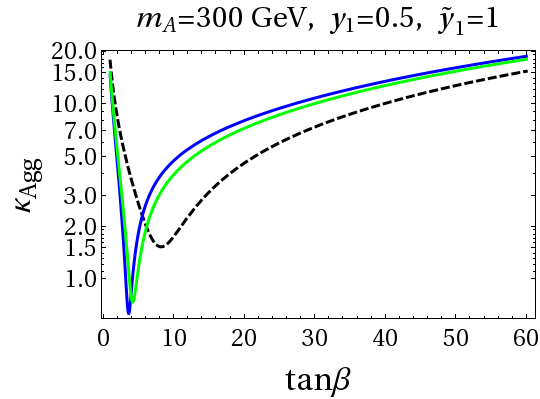}
\includegraphics[width=0.32\textwidth]{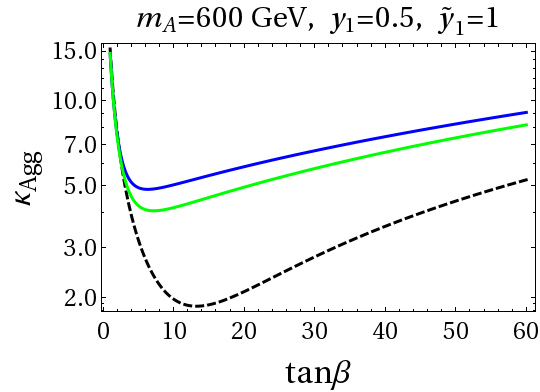}
\caption{ For the $MVQD_{11}$ model, $y^A_{11}$ (red), $y^h_{11}$ (blue) as a function of $\tan \beta$ (left); 
$\kappa_{A gg}$ as a function of $\tan \beta$ for $m_{A}=300~$GeV (middle) and $600~$GeV (right), with $y_{1} = 0.5$, $\tilde{y}_{1} =1$ and $M_{\psi}=800$~GeV (blue), $1000$~GeV (green)
.}
\label{sigmabr_c1}
\end{figure}

\begin{figure}[t]
\centering
\includegraphics[width=0.32\textwidth]{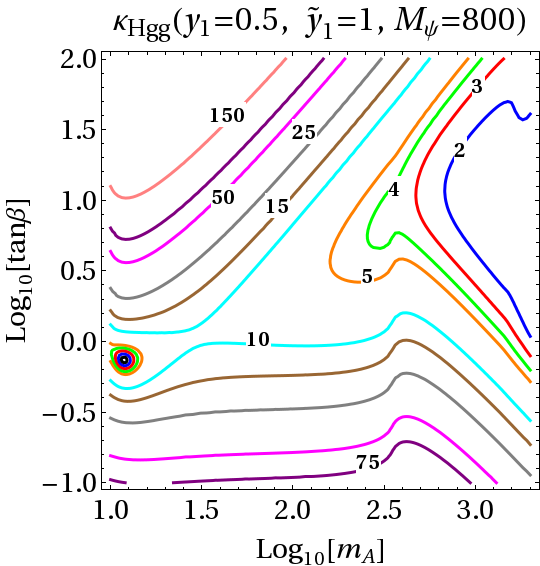}
\includegraphics[width=0.32\textwidth]{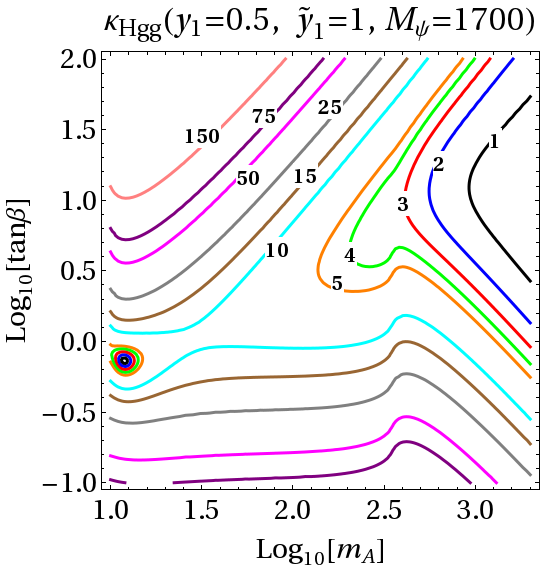}
\caption{Contours of $\kappa_{Hgg}$ for $y_{1} = 0.5$, $\tilde{y}_{1} =1$, for $M_{\psi} = M_{\chi}=$ 800 GeV (left), 1700 GeV (right) for $MVQD_{11}$~model. 
The corresponding contours in Type-II 2HDM are shown in Fig.~\ref{kphigg-2HDM-II.FIG}.}
\label{kappa_c1}
\end{figure}

In Fig.~\ref{effcouplingd}, we plot contours of $\kappa_{Agg}$ for $M_{\psi} = 800$~GeV, $1700$~GeV.
For comparison we have also plotted the corresponding contours in 2HDM-II.
Using this, one can read-off the $\sigma(gg\rightarrow A)$ at the 8~and~14~TeV LHC from Fig.~\ref{kgg_prod} in Sec.~\ref{ModInd.SEC}.
For comparison, the corresponding contours in the Type-II 2HDM (without the VLFs) are shown in Fig.~\ref{kphigg-2HDM-II.FIG}.
In Fig.~\ref{sigmabr_c1} (left) we plot $y^h_{11}$ and $y^A_{11}$ (defined in Eq.~(\ref{effcoupc1})) in the alignment limit ($\beta- \alpha=\pi/2$),
which shows that the $h$ couplings to the VLFs become very small as $\tan \beta$ increases.
Thus, the VLFs can modify $\sigma (g g \rightarrow A)$ and $\Gamma(A \rightarrow V V)$ significantly, while the $h$ remains SM-like as required by the LHC measurements of the $125$~GeV state. 
We find that the VLF contributions partially cancel the SM fermion contributions for a range of low  $\tan \beta$ values and for some ranges of the $m_{A}$,
while for larger $\tan \beta$ the effective couplings always increase compared to the 2HDM-II.
To illustrate this point more explicitly, we plot $\kappa_{Agg}$ as a function of $\tan \beta$ in Fig.~\ref{sigmabr_c1} for $m_{A}= 300 $~GeV and $600$~GeV.
The constraint on the 2HDM was nontrivial only for large $\tan\beta$ (see Fig.~\ref{constraints_tanb_2hdm}). 
Therefore, for large $\tan{\beta}$, since the $\kappa_{Agg}$ is bigger for this model compared to 2HDM (see Fig.~\ref{sigmabr_c1}), 
and the tree-level $\tau^{+} \tau^{-}$ BR from which the tightest constraint appears is almost unchanged, the constraint on this model will be tighter.
In Fig.~\ref{kappa_c1}, we plot contours of $\kappa_{Hgg}$ for $m_{A}= m_{H}$, in the alignment limit.
Corresponding contours in Type-II 2HDM are shown in Fig.~\ref{kphigg-2HDM-II.FIG}.
From this, one can also obtain $\sigma(gg \rightarrow H)$ from Fig.~\ref{kgg_prod}.

\medskip
\noindent \underline{\bf $MVQU_{22}$~model}: 
We introduce one doublet VLQ $(\psi)$ with hypercharge $Y_{\psi}$ and one singlet VLQ ($\xi$) with hypercharge $Y_{\psi}+1/2$, which couples only to $\Phi_{2}$. 
We add the following terms to the 2HDM-II Lagrangian
\begin{align}\label{case2b}
\mathcal{L}\supset \bar{\psi} i\slashed D \psi + \bar{\xi} i\slashed D  \xi -   {y_{2}}\bar \psi_{L} \tilde{\Phi}_{2} \xi_{R}  - \tilde{y}_{2}\bar \psi_{R} \tilde{\Phi}_{2} \xi_{L} +\text{h.c}
                    -M_{\psi} \bar \psi \psi -M_{\xi}\bar \xi \xi .
\end{align}
Here we do not include the terms $\bar \psi_{L} \tilde{\Phi}_{1} \xi_{R}$ and  $\bar \psi_{R} \tilde{\Phi}_{1} \xi_{L}$ as their effects have been considered in $MVQD_{11}$ model. 
As the BR($A \rightarrow VV)$s do not change much compared to the 2HDM-II case, we do not show them here.
Instead of the $\xi$ (with hypercharge $Y_{\psi}+1/2$) if we consider a VLF (say $\chi$) of hypercharge $(Y_{\psi}-1/2)$ we get a different model where the $\chi$ couples to the $\Phi_{2}$ 
instead of the $\tilde{\Phi}_{2}$. This will give similar effects to what we consider here. 

\begin{figure}[]
\centering
{}{\includegraphics[width=0.32\textwidth]{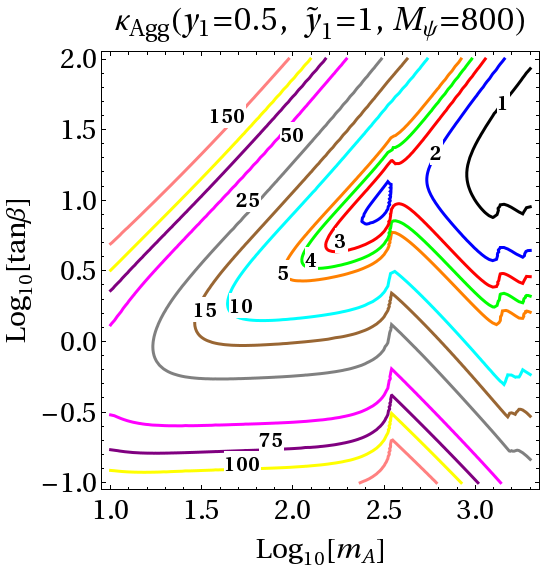}}\hspace{1em}%
{}{\includegraphics[width=0.32\textwidth]{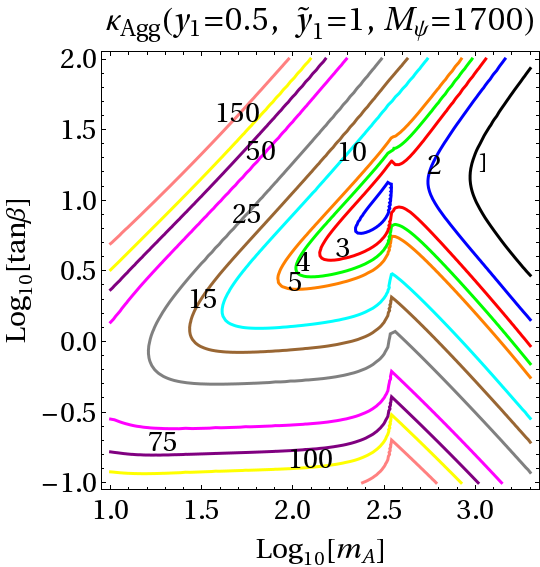}}\hspace{1em}%
\caption{Contours of $\kappa_{A gg }$ for $y_{2} = 0.5$, $\tilde{y}_{2} =1$, for $M_{\psi} = M_{\chi}=800$~GeV (left), 1700~GeV (right) for $MVQU_{22}$~model.}
\label{effcouplingu}
\end{figure}

\begin{figure}[t]
\centering
{}{\includegraphics[width=0.32\textwidth]{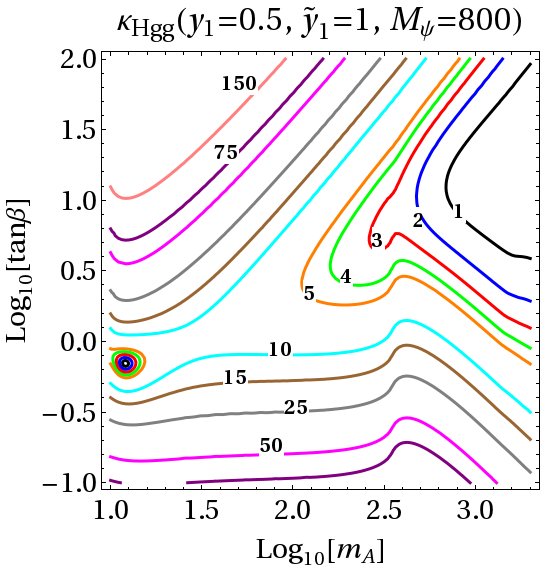}}\hspace{1em}%
{}{\includegraphics[width=0.32\textwidth]{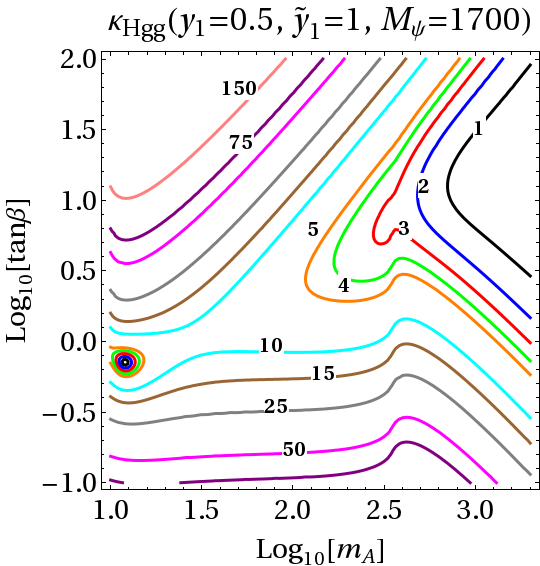}}\hspace{1em}%
\caption{Contours of $\kappa_{H gg }$ for $y_{2} = 0.5$, $\tilde{y}_{2} =1 $, for $M_{\psi} = M_{\chi}=800$~GeV (left), 1700~GeV (right) for $MVQU_{22}$~model.}
\label{kappa_c2}
\end{figure}

Similar to $MVQD_{11}$ model we diagonalize the mass matrix by an orthogonal rotation and define the couplings $y_{ij}^\phi$.
The mass eigenvalues, mixing angles and $y_{ij}^\phi$'s for this model can be found in App.~\ref{effcoup_c1_append}.
The effective couplings for this model are given in App.~\ref{kAvvGen.App}. 
As in $MVQD_{11}$~model, the $\kappa_{AVV}$ becomes zero when $y_{2}= \tilde{y}_{2}$. 
In Fig.~\ref{effcouplingu} we plot contours of $\kappa_{A gg }$ in $m_{A}$-$\tan \beta$ plane.
In the $MVQU_{22}$ model the VLF contributions to $\kappa_{Agg}$ are very small 
for $y_1 = 0.5$ and $\tilde y_1 = 1$, and therefore we do not show it explicitly.
This is particularly so for large $\tan\beta$ because $y_{ij}$'s are proportional to $\cos \beta$ which become small as 
$\tan \beta$ increases.
Similar conclusions hold for $\kappa_{Hgg}$.
In Fig.~\ref{kappa_c2} we plot $\kappa_{Hgg}$ using which one can read-off the $\sigma(gg \rightarrow H)$ from Fig.~\ref{kgg_prod} by reading $\kappa_{Agg}$ there as $\kappa_{Hgg}$ as mentioned earlier.
Since $\kappa_{Agg}$ and $\kappa_{Hgg}$ do not change much compared to the 2HDM-II, constraints on the $m_{A}$-$\tan \beta$ plane will almost remain same as in the 2HDM-II case.
Thus, VLFs if realized as in $MVQU_{22}$~model have little impact on the observables we consider here.

\medskip
\noindent \underline{\bf $MVQU_{12}$~model}: 
We introduce one doublet VLQ $(\psi)$ with hypercharge $Y_{\psi}$ and one singlet VLQ ($\xi$) with hypercharge $(Y_{\psi}+1/2)$.
We consider the case where $\xi_{R}$ couples only to $\Phi_{1}$ and $\xi_{L}$ couples only to $\Phi_{2}$.
To the 2HDM-II Lagrangian, we add 
\begin{align}\label{case3}
\mathcal{L}\supset \bar{\psi} i\slashed D \psi + \bar{\xi} i\slashed D  \xi -  ( {y_{1}}\bar \psi_{L} \tilde{\Phi}_{1} \xi_{R}  + \tilde{y}_{1}\bar \psi_{R} \tilde{\Phi}_{2} \xi_{L} +\text{h.c})
                    -M_{\psi} \bar \psi \psi -M_{\chi}\bar \xi \xi .
\end{align}
We get different models if instead of the couplings above, the $\psi_{R}$ couples to $\tilde{\Phi}_{1}$ and $\psi_{L}$ couples to $\tilde{\Phi}_{2}$, or, 
if instead of $\xi$ we introduce a VLF singlet (say $\chi$) with hypercharge $(Y_{\psi}-1/2)$ with couplings to $\Phi_{1}$ and $\Phi_{2}$. 
All these models have similar phenomenology as $MVQU_{12}$~model.
 
\begin{figure}[]
\centering
{}{\includegraphics[width=0.32\textwidth]{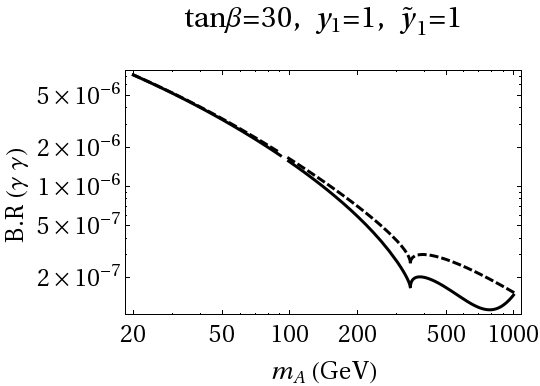}}
{}{\includegraphics[width=0.32\textwidth]{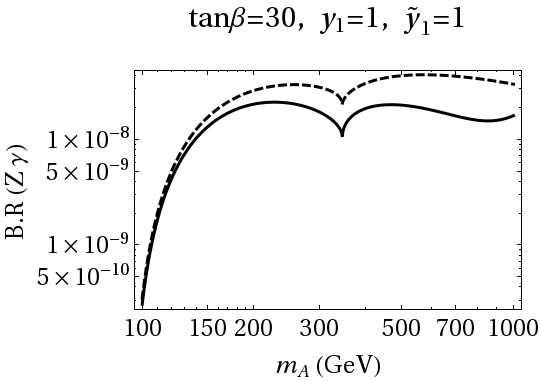}}
\caption{
BR($A \rightarrow \gamma \gamma$) (left) and BR($A \rightarrow Z \gamma$) (right) 
with $M_{\psi}= M_{\chi}=1000$~GeV for $\tan\beta=30$ for $MVQU_{12}$~model (solid-black),
and the corresponding variation in the Type-II 2HDM (dashed-black). 
These $BR$ for $\tan\beta=1$ and the $BR(A\to \tau\tau, bb, tt)$ are not explicitly shown here as they are identical to those in
Figs.~\ref{br_c1}~and~\ref{BRA2ff-2HDMII.FIG} respectively.
}
\label{br_c3}
\end{figure}

\begin{figure}[]
\centering
{}{\includegraphics[width=0.32\textwidth]{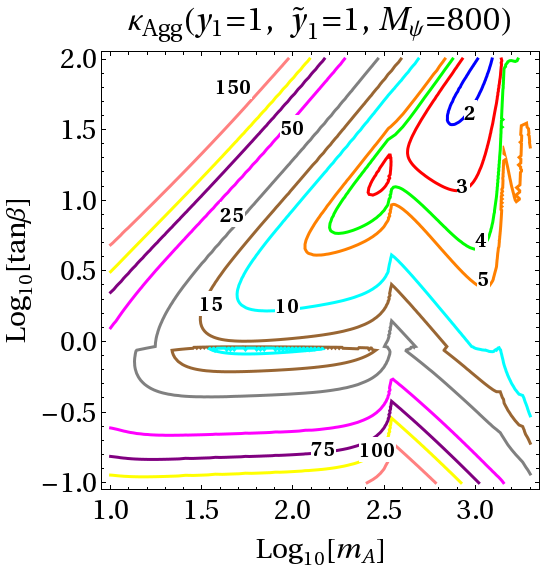}}\hspace{4em}%
{}{\includegraphics[width=0.32\textwidth]{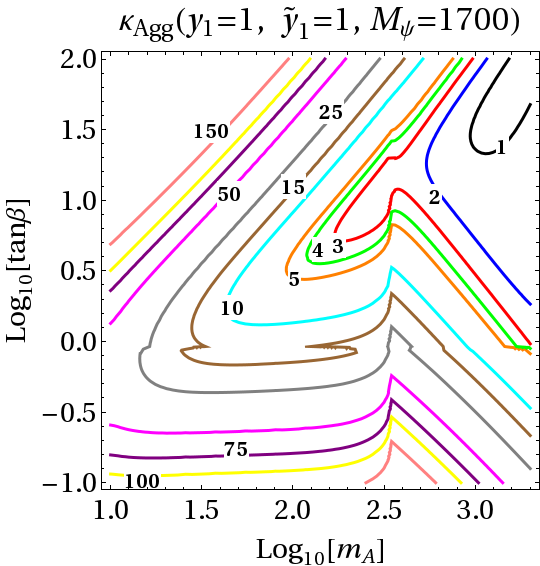}}
\caption{Contours of $\kappa_{A gg }$ for $y_{1} = 1$, $\tilde{y}_{1} =1 $, for $M_{\psi} = M_{\chi}=$ 800 GeV (left) and 1700 GeV (right) for $MVQU_{12}$~model.}
\label{effcouplingt}
\end{figure}

\begin{figure}[]
\centering
{}{\includegraphics[width=0.32\textwidth]{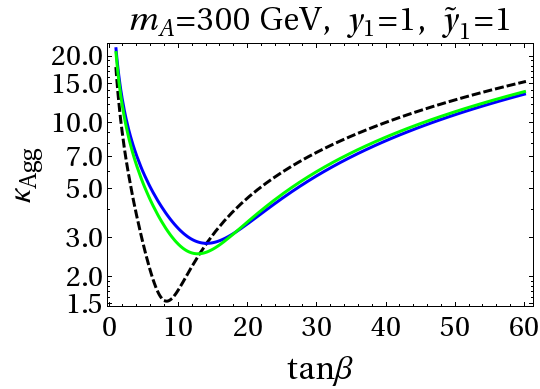}}\hspace{1em}%
{}{\includegraphics[width=0.32\textwidth]{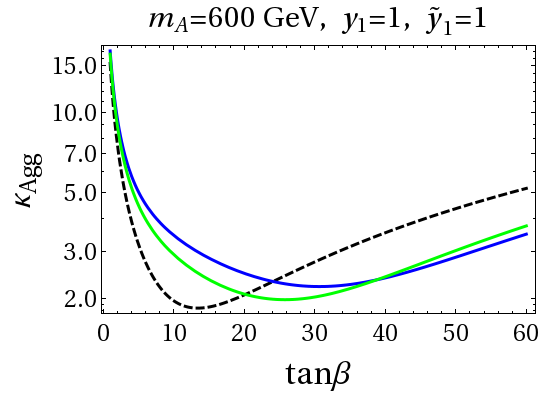}}
\caption{$\kappa_{A gg}$ with $\tan \beta$ for $m_{A}=300$~GeV (left) and $600$~GeV (right) with $y_{1} = 1$, $\tilde{y}_{1} =1 $ and $M_{\psi}=800$~GeV (blue), $1000$~GeV (green) for $MVQU_{12}$~model and 2HDM-II (dashed-black).}
\label{sigmabr_c3}
\end{figure}

\begin{figure}[]
\centering
{}{\includegraphics[width=0.32\textwidth]{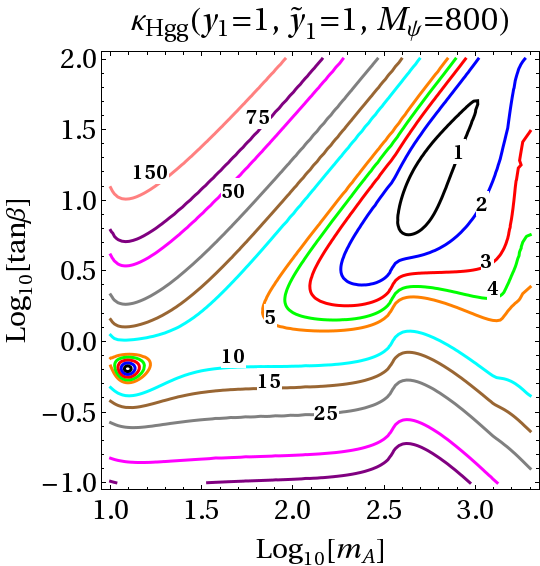}}\hspace{1em}%
{}{\includegraphics[width=0.32\textwidth]{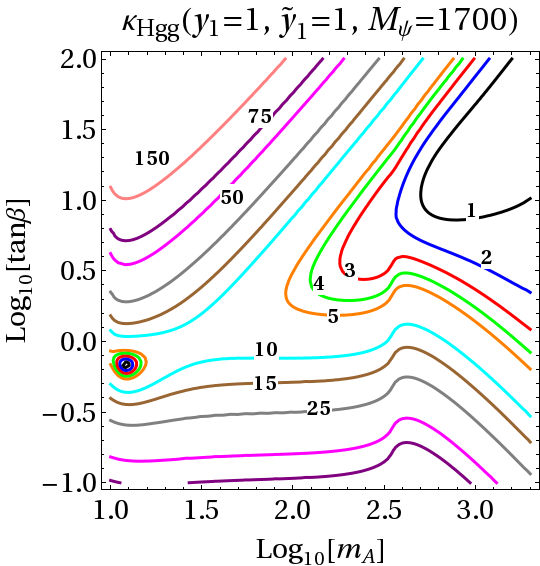}}\hspace{1em}%
\caption{ Contours of $\kappa_{H gg }$ for $y_{1} = 1$, $\tilde{y}_{1} =1 $, for $M_{\psi} = M_{\chi}=$ 800 GeV (left), 1700 GeV (right) for $MVQU_{12}$~model.}
\label{kappa_c3}
\end{figure}

The mass eigenvalues, mixing angles and $y_{ij}^\phi$'s for this model can be found in App.~\ref{effcoup_c3_append}.
The effective couplings for this model are given in App.~\ref{kAvvGen.App}. 
In this model, the effective couplings do not reduce to zero for $y_{1}= \tilde{y}_{1}$, unlike in $MVQD_{11}$~and~$MVQU_{22}$~models, 
as there are no additional $P$ and $C$ symmetries in the VLF sector.
In Fig.~\ref{br_c3}, we plot the BR($A\rightarrow V V$), BR($A\rightarrow b \bar b, \tau^{+} \tau^{-}, t \bar t$) for an example choice of $Y_{\psi}=1/6$.
The $BR(A\to \gamma\gamma, Z\gamma)$ for $\tan\beta = 1$, $y_1 = 0.5$, $\tilde y_1 = 1$ 
and the tree-level $BR(A\to \tau\tau, bb, tt)$ are not explicitly shown in Fig.~\ref{br_c3}
as they are identical to those shown for the $MVQD_{11}$ model in Fig.~\ref{br_c1}, 
and Type-II 2HDM in Fig.~\ref{BRA2ff-2HDMII.FIG} respectively.   
In Fig.~\ref{effcouplingt} we plot contours of $\kappa_{Agg}$ for $y_{1}=\tilde{y}_{1}=1$ and $M_{\psi}=M_{\xi}=800$~GeV and $1700$~GeV.
From this, one can obtain $\sigma(gg\rightarrow A)$ at the 8~and~14~TeV LHC from Fig.~\ref{kgg_prod} in Sec.~\ref{ModInd.SEC}.
For low values of $\tan \beta$ the effective coupling increases compared to the 2HDM-II case, while for larger values of $\tan \beta$ the effective coupling decreases compared to the 2HDM-II.
To show this more explicitly, we plot $\kappa_{Agg}$ with $\tan \beta$ in Fig.~\ref{sigmabr_c3}.
The decreased coupling is due to a destructive interference between the contributions from SM fermions and the VLFs.
If we reverse the sign of $y_{1}$ or $\tilde{y}_{1}$, we get the opposite effect; for low values of $\tan \beta $ the effective coupling decreases compared to the 
2HDM-II while for larger values of $\tan \beta$ the effective coupling increases compared to the 2HDM-II.
In Fig.~\ref{kappa_c3} we plot contours of 
$\kappa_{Hgg}$ in the alignment limit. 
From this, one can also obtain $\sigma(gg \rightarrow H)$ from Fig.~\ref{kgg_prod} by reading $\kappa_{Agg}$ there as $\kappa_{Hgg}$ as mentioned earlier.

\begin{figure}
\centering
{}{\includegraphics[width=0.32\textwidth]{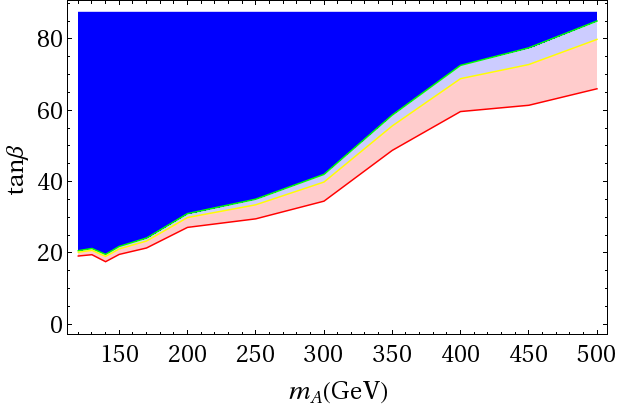}}\hspace{1em}%
{}{\includegraphics[width=0.32\textwidth]{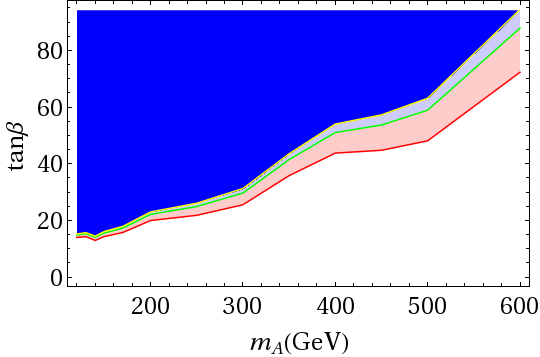}}
\caption{For $MVQU_{12}$~model, regions of the $m_{A}$-$\tan \beta$ parameter-space excluded at the $95 \%$ CL from $\phi \rightarrow \tau^{+} \tau^{-}$ decay when only $A$ is present (left), and when $A$ and $H$ are degenerate and both present (right), 
with $y_{1}=\tilde{y}_{1}=1$, $M_{\psi}=M_{\chi}= 800$~GeV (dark-blue region), 1000~GeV (light-blue and dark-blue regions). All shaded regions are excluded in the 2HDM-II.}
\label{ma_tanb_constraint}
\end{figure}

In Fig.~\ref{ma_tanb_constraint} we plot the region of the $m_{A}$-$\tan \beta$ parameter-space which is excluded at $95\,\%$ confidence level for two cases,
when only $A$ is present, and when $A$ and $H$ are degenerate and both present.
For comparison, we have also plotted the corresponding limit for the 2HDM-II case.
We see that the constraints are loosened compared to the 2HDM-II due to the presence of VLFs.
This happens because of the reduction of $\kappa_{Agg}$ ($\kappa_{Hgg}$) compared to the 2HDM-II.

Next, we add VLFs to the Type-X 2HDM and study the phenomenology of the neutral scalars. 

\subsubsection{Type-X 2HDM with VLQ-VLQ Yukawa couplings}

\begin{figure}
\centering
\includegraphics[width=0.32\textwidth]{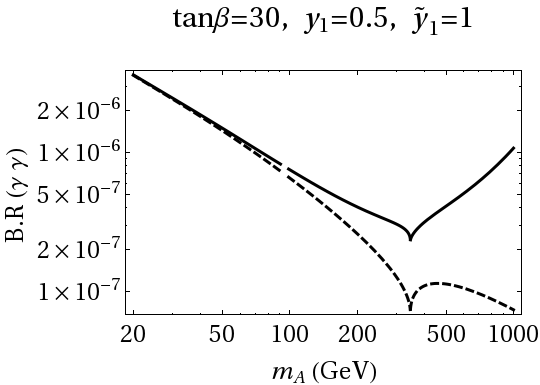}
\includegraphics[width=0.32\textwidth]{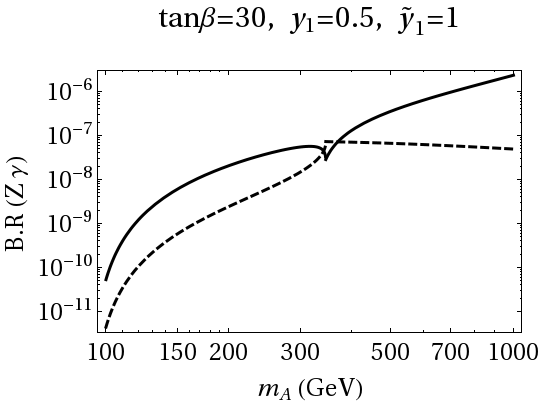}
\caption{
BR($A \to \gamma \gamma, Z \gamma $) with $M_{\psi}= M_{\chi}=1000$~GeV (solid-black) for $\tan \beta = 30$ 
for the $MVQDX_{11}$~model, and the corresponding variation in the 2HDM-X (dashed-black).} 
\label{br_m3}
\end{figure}

\medskip
\noindent \underline{\bf $MVQDX_{11}$~model}: 
To the 2HDM Type-X model in Eq.~(\ref{L2HDMX}), we introduce VLFs in a similar fashion as in $MVQD_{11}$~model 
as a representative case, and call it $MVQDX_{11}$~model. 
The other ways of coupling VLFs similar to $MVQU_{22}$~or~$MVQU_{12}$~model will be qualitatively similar to our results here.
We introduce a doublet VLQ $\psi=(\psi_{1}, \psi_{2})$ with hypercharge $Y_{\psi}$, and a singlet VLQ ($\chi$) 
with hypercharge $(Y_{\psi}-1/2)$ which couples only to $\Phi_{1}$.
To the 2HDM-X Lagrangian we add 
\begin{align}
\label{model3}
\mathcal{L}\supset \bar{\psi} i\slashed D \psi + \bar{\chi} i\slashed D  \chi -  ( {y_{1}}\bar \psi_{L} \Phi_{1} \chi_{R}  + \tilde{y}_{1}\bar \psi_{R} \Phi_{1} \chi_{L} +\text{h.c})
                    -M_{\psi} \bar \psi \psi -M_{\chi}\bar \chi \chi .
\end{align}
The effective couplings of $A$ with VLFs are same as in $MVQD_{11}$~model and can be read off from App.~\ref{kAvvGen.App}.
In Fig.~\ref{br_m3} we show BR($A \rightarrow VV$) including the VLF contributions for the $MVQDX_{11}$~model, 
and the tree-level BR($A \rightarrow \tau^{+} \tau^{-}, b \bar b, t \bar t$) is unchanged from what are shown in 
Fig.~\ref{BRA2ff-2HDM-X.FIG}.
BR($A \to \gamma \gamma, Z \gamma $) for $\tan\beta = 1$, $y_1=0.5$, $\tilde y_1=1$ are almost identical to the 2HDM values 
shown in Fig.~\ref{br_c1} and are therefore not shown explicitly in Fig.~\ref{br_m3}.
For $\tan \beta =30$, BR($A \rightarrow \gamma \gamma, Z \gamma$) is increased compared to 2HDM-II, 
because for large $\tan \beta$, $\Gamma(A \rightarrow b \bar b)$ becomes much smaller in 2HDM-X.

\begin{figure}[]
\centering
\includegraphics[width=0.32\textwidth]{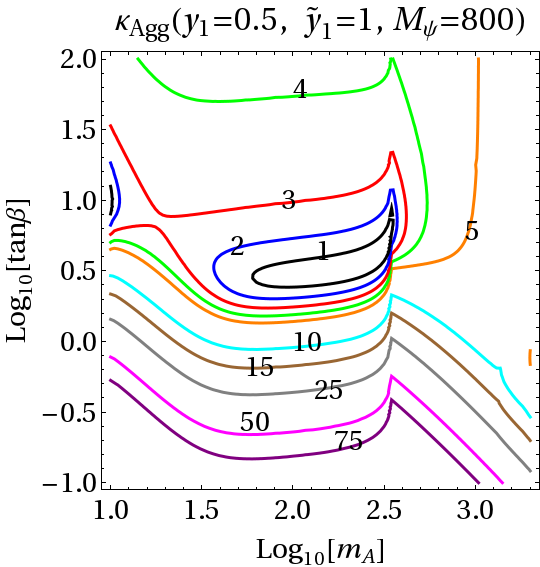}
\includegraphics[width=0.32\textwidth]{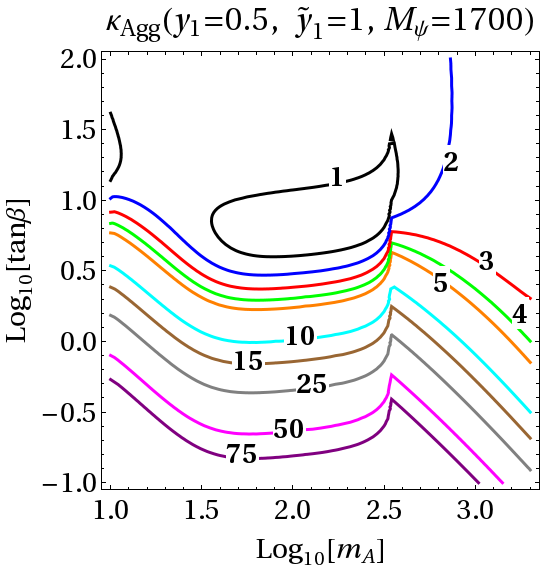}
\caption{Contours of $\kappa_{A gg }$ for $y_{1} = 0.5$, $\tilde{y}_{1} =1 $, 
for $M_{\psi} = M_{\chi}=$ 800 GeV (left), 1700 GeV (right) for $MVQDX_{11}$~model. 
The corresponding contours in Type-X 2HDM is shown in Fig.~\ref{kphigg-2HDM-X.FIG}.}
\label{effcouplingm3}
\end{figure}

\begin{figure}[t]
\centering
  {}{\includegraphics[width=0.32\textwidth]{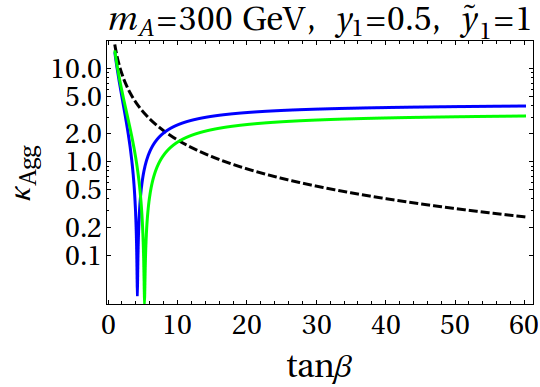}}\hspace{1em}%
  {}{\includegraphics[width=0.32\textwidth]{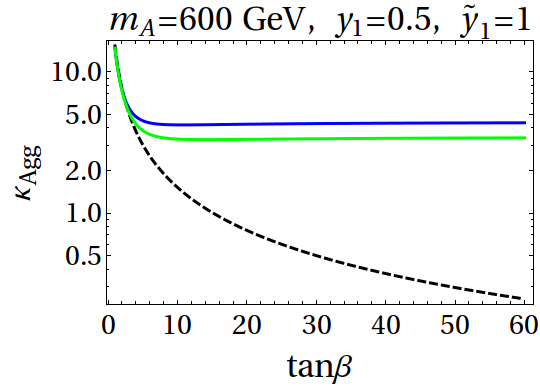}}
  \caption{ $\kappa_{A gg}$ with $\tan \beta$ for $m_{A}=300$~GeV (left) and $600$~GeV (right) with $y_{1} = 0.5$, $\tilde{y}_{1} =1 $ and $M_{\psi}=800$~GeV (blue), $1000$~GeV (green) for $MVQDX_{11}$~model and 2HDM-X (dashed-black).}
  \label{sigmabr_m3}
\end{figure}

\begin{figure}[t]
\centering
\includegraphics[width=0.32\textwidth]{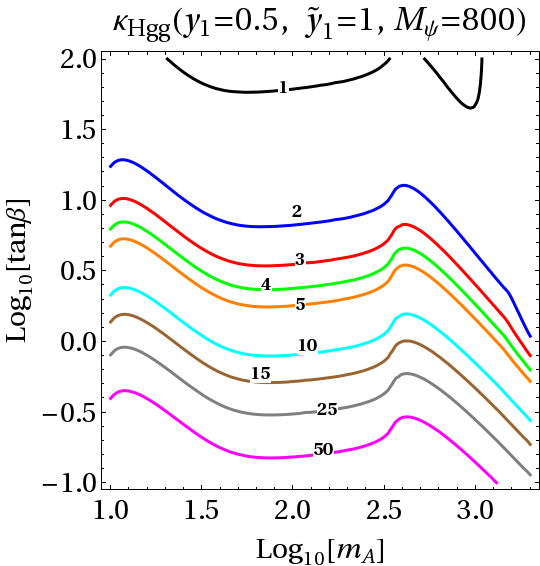}
\includegraphics[width=0.32\textwidth]{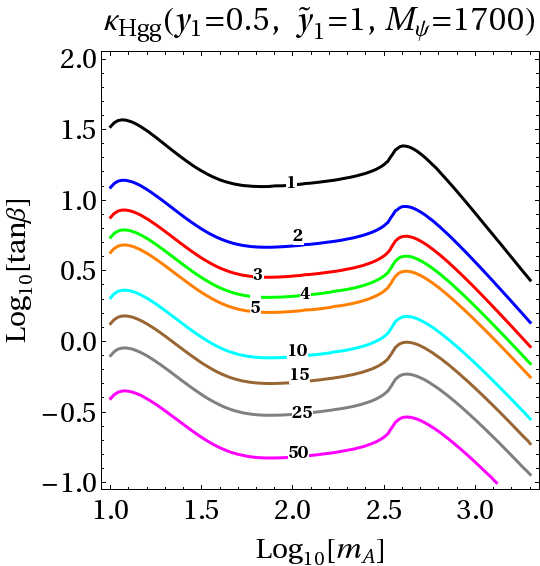}
\caption{ Contours of $\kappa_{H gg }$ for $y_{1} = 0.5$, $\tilde{y}_{1} =1 $, for $M_{\psi} = M_{\chi}=$~800 GeV (left), 1700 GeV (right) for $MVQDX_{11}$~model. The corresponding contours in Type-X 2HDM is shown in Fig.~\ref{kphigg-2HDM-X.FIG}.}
\label{kappa_m3}
\end{figure}

In Fig.~\ref{effcouplingm3} we plot contours of $\kappa_{Agg}$.
The $\kappa_{\phi gg}$ contours in 2HDM-X (without VLFs) are shown in Fig.~\ref{kphigg-2HDM-X.FIG}.
Using these plots, one can read off $\sigma(gg \rightarrow A)$ for 8~TeV and 14~TeV LHC from Fig.~\ref{kgg_prod} in Sec.~\ref{ModInd.SEC}. 
As expected, for large $\tan \beta$, $\kappa_{Agg}$ is significantly larger in this model compared to 2HDM-X since the VLFs contribute substantially while the SM quark contributions alone are very small.
In order to show explicitly how large the change is, we plot $\kappa_{Agg}$ as a function of $\tan \beta$ 
for $m_{A}=300$~GeV and $600$~GeV in Fig.~\ref{sigmabr_m3}.
In Fig.~\ref{kappa_m3} we plot contours of $\kappa_{Hgg}$ in $m_{A}$--$\tan \beta$ plane in the alignment limit. 
From this, one can also obtain $\sigma(gg \rightarrow H)$ from Fig.~\ref{kgg_prod}.
\subsubsection{Type-II 2HDM with VLL-VLL Yukawa couplings}
%

\begin{figure}
\centering
{}{\includegraphics[width=0.32\textwidth]{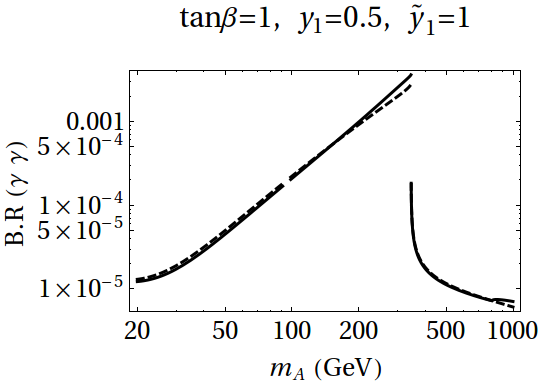}}\hspace{1em}%
{}{\includegraphics[width=0.32\textwidth]{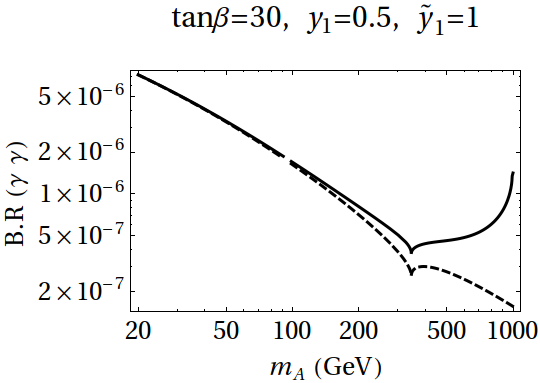}}
\caption{BR($A \rightarrow \gamma \gamma$) with $M_{\psi}= M_{\chi}=$ 500 GeV, $y_{1}=0.5$, $\tilde{y_{1}}=1$ for $\tan \beta =1$ (left) and $\tan \beta =30$ (right) for $MVLE_{11}$.} 
\label{br_vll}
\end{figure}

\medskip
\noindent \underline{\bf $MVLE_{11}$ model}:
Vector-like leptons do not contribute in $g g \rightarrow A$, but can contribute in $A \rightarrow \gamma\gamma, Z\gamma$.
We show the effect of VLLs in a simple model similar to $MVQD_{11}$~model, but with VLLs instead of VLQs. 
We introduce one doublet VLL ($\psi$) with hypercharge, $Y_{\psi}$ and one singlet VLL ($\chi$) with hypercharge, $Y_{\psi}-1/2$.
The Lagrangian we consider is exactly the same as in Eq.~(\ref{case2a}), except here the VLLs $\psi$ and $\chi$ do not couple to gluons.
The effective couplings are the same as for $MVQD_{11}$~model except for color factors.
As an example, we choose $Y_{\psi}=-1/2$ and plot BR($A \rightarrow \gamma \gamma$) as a function of $m_{A}$ in Fig.~\ref{br_vll}, with $M_{\psi}= M_{\chi}$= 500 GeV, for $\tan \beta=1$ and $30$.
We see that the effect of VLLs is qualitatively similar to vector-like quarks; for low $\tan \beta$ the effect of VLLs is negligible
while for large $\tan \beta$ and large $m_{A}$ VLL contributions are significant. 
Near $m_{A}=1000$~GeV, the VLL contribution is quite large due to them going onshell for our choice of VLL mass of $500$~GeV. 
BR($A \rightarrow Z \gamma$) will show the same behavior.

\section{Conclusions}
\label{Concl.SEC}

Many theories beyond the standard model (BSM) contain new CP-odd and CP-even neutral scalars $\phi = \{A,H$\} 
and new vector-like fermions ($\psi_{VL}$).
We study the LHC phenomenology of $\phi$ taking into account $\psi_{VL}$ contributions to 
$\phi gg$, $\phi\gamma\gamma$ and $\phi Z \gamma$ couplings at the one-loop level. 

In Sec.~\ref{ModInd.SEC} we write an effective Lagrangian with $\phi$ coupled to standard model (SM) gauge-bosons and fermions. 
We focus only on $\phi$ Yukawa couplings to third generation SM fermions, namely $t,b,\tau$, since these are usually the bigger couplings in most BSM extensions. 
The couplings of the $A$ to standard model $W,Z$ gauge bosons (i.e. $AVV$ couplings) cannot occur from renormalizable operators in a CP-conserving sector, 
but can be induced as loop-generated non-renormalizable operators. 
These operators are induced by SM fermions and also the heavy $\psi_{VL}$.  
In Sec.~\ref{ModInd.SEC} we present model-independent results that are useful whatever be the origin of these effective couplings. 
In Fig.~\ref{kgg_prod} we present the 8~TeV and 14~TeV LHC $gg\to \phi$ (gluon-fusion channel) cross-sections as a function of the effective couplings. 
We also obtain limits on the effective couplings from the 8~TeV LHC data on the $\gamma \gamma$, $\tau^{+} \tau^{-}$ and $t \bar t$ modes. 
We do not include the $b\bar b$ decay mode and the $b$-quark associated production channels in this work.

We define some simple models in Sec.~\ref{Models.SEC} that are representative of BSM constructions as far as the phenomenology of $\phi$ is concerned. 
These models include $\phi$ and $\psi_{VL}$ in the singlet and doublet representations of $SU(2)$. 
In the doublet case, we focus on the two-Higgs-doublet (2HDM) Type-II and Type-X models.  
We compute the $\phi gg$, $\phi\gamma\gamma$ and $\phi Z \gamma$ effective couplings induced by the SM fermions and vector-like fermions at the 1-loop level and present analytical expressions for them in App.~\ref{effCoups.APP}. 
For the various models we define, we present the effective couplings $\kappa_{\phi gg}$, $\kappa_{\phi \gamma\gamma}$, $\kappa_{\phi Z \gamma}$, and, 
$BR(A\to \gamma\gamma, Z\gamma)$ and $BR(A\to f\bar f)$ for $f=\{ \tau, b, t\}$ 
as a function of the model parameters. 
From the $\kappa_{\phi gg}$ and the $BR$ into one of these modes, one can see if a point in parameter-space in a given model is allowed by the 8~TeV data from our plots in
Sec.~\ref{ModInd.SEC}. 
One can also read-off the gluon-fusion cross-section at the 8~TeV and 14~TeV LHC from Fig~\ref{kgg_prod}. 
Interestingly, for some of the 2HDM cases we studied, we find that the addition of vector-like fermions loosens the constraint compared to the 2HDM alone, and allows more of the parameter-space. 
This can be seen for instance in Fig.~\ref{ma_tanb_constraint}. 
The 14~TeV LHC gluon-fusion c.s. of the $\phi$ and its BRs in the different models we present should be useful in identifying 
allowed regions of parameter space and promising discovery channels of the $\phi$.  
In this context, it is interesting to explore the possibility of the $\phi$ being the state responsible for 
the recent 750~GeV excess of diphoton events at the 13~TeV LHC~\cite{ATLAS-750GeVExcess-2015,CMS:2015dxe};
this is the subject of a separate work.

\medskip
\noindent {\it Acknowledgments:}
We thank Gautam Bhattacharya and Dipankar Das for a discussion on the unitarity bounds in 2HDM, and V.~Ravindran for a discussion on $b\bar b$ fusion. 

\setcounter{section}{0}
\renewcommand\thesection{\Alph{section}}               
\renewcommand\thesubsection{\Alph{section}.\arabic{subsection}}
\renewcommand\thesubsubsection{\Alph{section}.\arabic{subsection}.\arabic{subsubsection}}

\renewcommand{\theequation}{\Alph{section}.\arabic{equation}}    
\renewcommand{\thetable}{\Alph{section}.\arabic{table}}          
\renewcommand{\thefigure}{\Alph{section}.\arabic{figure}}        

\section{Couplings, masses and mixing angles in various models}
\label{effCoups.APP}
Here we provide the explicit expressions for the mixing angles and the mass eigenvalues in the different models we have defined in Sec.~\ref{Models.SEC}.
We also provide expressions for the $\kappa_{ij}$'s, $y_{ij}$'s defined in Eq.~(\ref{effcoupc1}).
Sec.~\ref{tpbpMangle.App} contains the mixing angles and the mass eigenvalues for $MVU$, $MVD$ and $MVQ$ models.
Sec.~\ref{effcoup_c1_append} contains explicit expressions for $y_{ij}$'s and $\kappa_{ij}$'s 
for $MVQD_{11}$, $MVQU_{22}$ and $MVQDX_{11}$ models.
Sec.~\ref{effcoup_c3_append} contains explicit expressions for $y_{ij}$'s and $\kappa_{ij}$'s 
for $MVQU_{12}$ model.
In what follows we will use the notations $c_{L,R} = \cos \theta_{L,R}$, $s_{L,R} = \sin \theta_{L,R}$, $c_\beta = \cos \beta$ and $s_\beta = \sin \beta$.

\subsection{$MVU$, $MVD$, $MVQ$ models}
\label{tpbpMangle.App}
In this section we give the mixing angles and the mass eigenvalues for $MVU$, $MVD$ and $MVQ$ models.
The mixing angles $\theta^U_{L,R}$, for $MVU$ model are given by,
\begin{align}
\label{tpMangle}
& \tan 2 \theta^U_L = \frac{2 \sqrt{2} y_1 v_1 M_{\psi}}{y_u^2 v_2^2 - 2 M_{\psi}^{2}  + y_1^{2} v_1^{2}}, \hspace{2mm}
 \tan 2 \theta^U_R = \frac{2 \sqrt{2} y_1 y_u v_1 v_2 }{y_u^2 v_2^2 - 2 M_{\psi}^{2}  - y_1^{2} v_1^{2}}. 
\end{align}
The mass eigenvalues for the EM charge-2/3 fermions in $MVU$ model are given by,
\begin{align}
\label{tpMEvalue}
 m_{t,t_2} = \frac{1}{2}\left(\sqrt{\left(\frac{y_u}{\sqrt{2}} v_2 + M_{\psi} \right)^2 + \frac{y_1^2}{2} v_1^2 } \mp \sqrt{\left(\frac{y_u}{\sqrt{2}} v_2 - M_{\psi} \right)^2+ \frac{y_1^2}{2} v_1^2 }\right)
\end{align}
The mixing angle and mass eigenvalues for $MVD$ model 
are obtained from Eq.~(\ref{tpMangle}) and Eq.~(\ref{tpMEvalue}) by the replacements $y_1 \rightarrow y_2$,
$y_u v_2\rightarrow y_d v_1$ and $M_\psi \rightarrow M_\chi$.
The mixing angles $\theta^U_{L,R}$, for $MVQ$ model are given by,
\begin{align*}
& \tan 2 \theta^U_R = \frac{2 \sqrt{2} \tilde{y}_1^{eff} v_2 M_{Q}^{eff}}{ 2 \left(M_{Q}^{eff} \right)^{2} -\left(y_u^{eff}\right)^2 v_2^2  + (\tilde{y}_1^{eff})^{2} v_2^{2}}, \hspace{2mm}
 \tan 2 \theta^U_L = \frac{2 \sqrt{2} \tilde{y}_1^{eff} y_u^{eff} v_2^2 }{ 2 \left(M_{Q}^{eff}\right)^{2}-\left(y_u^{eff} \right)^2 v_2^2  - \left(\tilde{y}_1^{eff}\right)^{2} v_2^{2}}. 
\end{align*}
The mass eigenvalues for the EM charge-2/3 fermions in $MVQ$ model are given by,
\begin{align*}
 m_{t,t_2} = \frac{1}{2}\left(\sqrt{\left(\frac{y_u^{eff}}{\sqrt{2}} v_2 + M_{Q}^{eff} \right)^2 + \frac{\left( \tilde{y}_1^{eff} \right)^2}{2} v_2^2 } \mp \sqrt{\left(\frac{y_u^{eff}}{\sqrt{2}} v_2 - M_{Q}^{eff} \right)^2+ \frac{\left( \tilde{y}_1^{eff} \right)^2}{2} v_2^2 }\right).
\end{align*}

\subsection{$MVQD_{11}$, $MVQU_{22}$ models}
\label{effcoup_c1_append}
In this section we give the expressions for the $y_{ij}$'s and $\kappa_{ij}$'s for $MVQD_{11}$, $MVQU_{22}$ models.
The couplings $\kappa_{ij}$ defined in Eq.~(\ref{effcoupc1}) for $MVQD_{11}$, $MVQU_{22}$ models and also for $MVQDX_{11}$~model are given by,
$ \kappa_{11} = (g/c_{W}) [ (T^3/2) (c^{2}_{L} +c^{2}_{R}) -  Q s_{W}^{2} ]$,
$\kappa_{22} = (g/c_{W}) [  (T^3/2) (s^{2}_{L} +s^{2}_{R}) - Q s_{W}^{2} ]$, 
$ \kappa_{12} =  - (g/c_{W}) (T^3/2) (s_L c_L + s_R c_R).$
The mass eigenvalues $M_{1,2}$ (in Eq.~\ref{effcoupc1}) for $MVQD_{11}$ model are given by,
 \begin{align}
 \label{MeigenVal.D11}
 &M_{1,2} =      \frac{1}{2} \sqrt{(M_{\psi} + M_{\chi})^{2} +\frac{1}{2} c_{\beta}^{2} v^{2} (y_{1}-\tilde{y}_{1})^{2}} \pm \sqrt{(M_{\psi} - M_{\chi})^{2} + 
  \frac{1}{2}v^{2} c_{\beta}^{2}(y_{1}+ \tilde{y}_{1})^{2}}
 \end{align}
and the mixing angles $\theta_{L,R}$ for $MVQD_{11}$ model are given by,
 \begin{align}
  \label{Mangel.D11}
 &\tan 2\theta_{L} = \frac{2\sqrt{2} v c_{\beta} (y_{1} M_{\chi} + \tilde{y}_{1} M_{\psi})}{2(M_{\psi}^{2} - M_{\chi}^{2}) - v^{2} s_{\beta}^{2}(\tilde{y}_{1}^{2} -y_{1}^{2})},~
&\tan 2\theta_{R} = \frac{2\sqrt{2} v c_{\beta} (y_{1} M_{\chi} + \tilde{y}_{1} M_{\psi})}{2(M_{\psi}^{2} - M_{\chi}^{2}) + v^{2} s_{\beta}^{2}(\tilde{y}_{1}^{2} -y_{1}^{2})}.
\end{align}
The mass eigenvalues and the mixing angles for $MVQU_{22}$~model can be obtained from Eq.~(\ref{MeigenVal.D11}) and Eq.~(\ref{Mangel.D11})
by the replacements $y_1 \to y_2$ and $c_\beta \to s_\beta$.
The couplings $y^A_{ij}$'s (in Eq.~\ref{effcoupc1}) for $MVQD_{11}$~model are given by,
$y^A_{11} =  (1/\sqrt{2}) s_{\beta} (-y_{1} c_{L} s_{R} +\tilde{y}_{1} s_{L} c_{R})$,
$y^A_{22} =  (1/\sqrt{2})  s_{\beta} (y_{1} s_{L} c_{R} - \tilde{y}_{1} c_{L} s_{R})$,
$y^A_{12} =  -(1/\sqrt{2}) s_{\beta} (y_{1} c_{L} c_{R} + \tilde{y}_{1} s_{L} s_{R})$,
$y^A_{21} =  (1/\sqrt{2}) s_{\beta}  (y_{1} s_{L} s_{R} + \tilde{y}_{1} c_{L} c_{R})$.
The $y^{A}_{ij}$'s in $MVQU_{22}$~model can be obtained from the $y^{A}_{ij}$'s in $MVQD_{11}$~model by the replacements $y_1 \to y_2$ and $s_\beta \to c_\beta$.
The couplings $y^h_{ij}$ (in Eq.~\ref{effcoupc1}) are given by,
 $y^h_{11} = -(1/\sqrt{2}) s_{\alpha} (y_{1} c_{L} s_{R} + \tilde{y}_{1} s_{L} c_{R})$,
 $y^h_{22} = (1/\sqrt{2})  s_{\alpha} (y_{1} s_{L} c_{R} + \tilde{y}_{1} c_{L} s_{R})$,
 $y^h_{12} = -(1/\sqrt{2}) s_{\alpha} (y_{1} c_{L} c_{R} - \tilde{y}_{1} s_{L} c_{R})$,
 $y^h_{21} = -(1/\sqrt{2}) s_{\alpha}  (-y_{1} s_{L} s_{R}  + \tilde{y}_{1} c_{L} c_{R})$.
The $y^h_{ij}$'s in $MVQU_{22}$~model can be obtained from $y^h_{ij}$'s in $MVQD_{11}$ model by the replacements  $y_1 \to y_2$ and $s_\alpha \to - c_\alpha$.
The couplings $y^H_{ij}$ (in Eq.~\ref{effcoupc1}) can be obtained from the $y^h_{ij}$'s in $MVQD_{11}$ model
by the replacements, $s_\alpha \to -c_\alpha$ in case of $MVQD_{11}$
and $s_\alpha \to -s_\alpha $ for $MVQU_{22}$~model.


\subsection{$MVQU_{12}$~model}
\label{effcoup_c3_append}

In this section we give the expressions for the $y_{ij}$'s and $\kappa_{ij}$'s for the $MVQU_{12}$ model.
The couplings $\kappa_{ij}$ for $MVQU_{12}$~model are same as in $MVQD_{11}$~model.
The mass eigenvalues are given by
\begin{align}
 &M_{1,2} = \frac{1}{2} \sqrt{(M_{\psi} + M_{\xi})^{2} +\frac{1}{2}  v^{2} (y_{1} c_{\beta}-\tilde{y}_{1} s_{\beta})^{2}} \pm \sqrt{(M_{\psi} - M_{\xi})^{2} + 
 \frac{1}{2}v^{2} (y_{1} c_{\beta} + \tilde{y}_{1} s_{\beta})^{2}}.
\end{align}
 and the mixing angles $\theta_{L,R}$ are given by
 \begin{align}
 &\tan 2\theta_{L} = \frac{2\sqrt{2} v (y_{1}  c_{\beta} M_{\xi} + \tilde{y}_{1} s_{\beta}  M_{\psi})}{2(M_{\psi}^{2} - M_{\xi}^{2}) - v^{2} (\tilde{y}_{1}^{2} s_{\beta}^{2} -y_{1}^{2} c_{\beta}^{2})},
\hspace{0.2cm}
  &\tan 2\theta_{R} = \frac{2\sqrt{2} v (y_{1} c_{\beta} M_{\xi} + \tilde{y}_{1} s_{\beta} M_{\psi})}{2(M_{\psi}^{2} - M_{\xi}^{2}) + v^{2} (\tilde{y}_{1}^{2} s_{\beta}^{2} -y_{1}^{2} c_{\beta}^{2})}.
\end{align}
 The couplings $y^A_{ij}$ are given by,
 $y^A_{11} =       (1/\sqrt{2})  (y_{1} s_{\beta} c_{L} s_{R} + \tilde{y}_{1} c_{\beta} s_{L} c_{R})$,
 $y^A_{22} =      -(1/\sqrt{2})  (y_{1} s_{\beta} s_{L} c_{R} + \tilde{y}_{1} c_{\beta} c_{L} s_{R})$,
 $y^A_{12} =      (1/\sqrt{2})  (y_{1} s_{\beta} c_{L} c_{R} - \tilde{y}_{1}  c_{\beta} s_{L} s_{R})$,
 $y^A_{21} =      -(1/\sqrt{2}) (y_{1} s_{\beta} s_{L} s_{R} - \tilde{y}_{1} c_{\beta} c_{L} c_{R})$.
 The couplings $y^h_{ij}$ are given by,
 $y^h_{11} =       (1/\sqrt{2})  (-y_{1} s_{\alpha} c_{L} s_{R} + \tilde{y}_{1} c_{\alpha} s_{L} c_{R})$,
 $y^h_{22} =      (1/\sqrt{2})  (y_{1} s_{\alpha} s_{L} c_{R} - \tilde{y}_{1} c_{\alpha} c_{L} s_{R})$,
 $y^h_{12} =      -(1/\sqrt{2})  (y_{1} s_{\alpha} c_{L} c_{R} + \tilde{y}_{1}  c_{\alpha} s_{L} s_{R})$,
 $y^h_{21} =      (1/\sqrt{2}) (y_{1} s_{\alpha} s_{L} s_{R} + \tilde{y}_{1} c_{\alpha} c_{L} c_{R})$.
The $y_{ij}^H$'s can be obtained from $y^h_{ij}$'s by the replacements $s_\alpha \to -c_\alpha$ and $c_\alpha \to s_\alpha$.
\section{The $\kappa_{\phi gg,\, \phi\gamma\gamma,\, \phi Z\gamma,\, A WW,\, A ZZ}$ effective couplings in various models}
\label{kAvvGen.App}

In this section we give the expressions for the $\kappa_{ \{ \phi gg,\, \phi \gamma\gamma,\, \phi Z\gamma,\, A ZZ,\, A WW \}}$ in the various models we have considered in Sec.~\ref{Models.SEC}.

\medskip
\noindent\underline{$\kappa_{\phi gg},~\kappa_{\phi \gamma\gamma}$}:
The 1-loop expressions for the $\phi g g$ and $\phi \gamma\gamma$ amplitudes $\kappa_{\phi gg}$ and $\kappa_{\phi \gamma\gamma}$ respectively,
with $\phi = \{h,H,A\}$ are given here. 
These amplitudes are induced by quarks whose effective Lagrangian can be written as ${\cal L}^{f}_{\phi} \supset m_{f} \bar{f} f + y_{\phi ff} \phi \bar{f} f$.  
Defining $r_{f} = m_{f}^{2}/m^{2}_{\phi}$ and with $f$ running over all colored fermion species with mass $m_{f}$ and Yukawa couplings $y_{\phi ff}$, 
and with the electric charge of the fermion ($f$) denoted by $Q_{f}$,
the general expressions for $\kappa_{\phi gg}$ and $\kappa_{\phi \gamma\gamma}$ are given as
\bea
\label{kphiag}
\kappa_{\phi \gamma \gamma} = 2 e^2 \sum_{f} N_{c}^f Q_{f}^{2} \, y_{\phi ff} \frac{M}{m_f}  F_{1/2}^{(1)}(r_{f}) \ , \qquad
\kappa_{\phi gg} =  g_{s}^{2} \sum_{f} y_{\phi ff} \frac{M}{m_f}  F_{1/2}^{(1)}(r_{f})  \ , \\
{\rm\text with} \ \ 
F_{1/2}^{(1)}(r_{f}) = 4 r_{f} \left(\int_{0}^{1}dy \int_{0}^{1-y} dx \frac{g(x,y)}{(r_{f}-x y)}\right) \ , \nonumber
\eea
and $g(x,y)= (1 - 4 x y)$ for the CP-even scalars ($h,H$) and $1$ for the CP-odd scalar ($A$).
Here $M$ is a mass scale defined in Eq.~(\ref{LEff.EQ}), which we set to $1~$TeV for numerical results.
Compared to $\kappa_{\phi \gamma \gamma}$, $\kappa_{\phi gg}$ has an extra factor of 1/2 which compensates 
for our definition of $\Gamma(\phi \to gg)$ in Eq.~(\ref{GmA2XX.EQ}) with a relative factor of eight compared to 
$\Gamma(\phi \to \gamma \gamma)$ while the actual color factor is really two.
The expressions for $F^{(1)}_{1/2}$ in Eq.~(\ref{kphiag}) match with the closed form expressions given in Ref.~\cite{Gunion:1989we}.

\medskip
\noindent\underline{$\kappa_{A Z \gamma}$}:
Here we give the general expressions for $\kappa_{AZ\gamma}$ (defined in Eq.~\ref{LEff.EQ}) for the different models we have considered.
For $SVU$ and $SVQ$ models,
\begin{align*}
\kappa_{A Z \gamma}=2  e \frac{g}{c_{W}} \sum_{i}   N_{c}^i Q_{i} (T^{i}_{3} - Q_{i}s_{W}^{2}) y_{A} \frac{M}{m_i} F_{1/2}^{(2)}(r_i, r_Z) \ , \\
{\rm\text with} \ \
F_{1/2}^{(2)}(r_i, r_Z) = 4 r_{i} \int_{0}^{1}dy \int_{0}^{1-y} dx \frac{ 1 }{r_{i} + (r_{Z} -1) x y + r_{Z}(x^{2}-x)} \ . \nonumber 
\end{align*}
For $SVU$ model only one VLF contributes to $\kappa_{AZ \gamma}$.
For $MVQD_{11}$, $MVQU_{22}$, $MVQU_{12}$ and $MVQDX_{11}$ models $\kappa_{A Z \gamma } = \kappa_{A Z\gamma}^1 + \kappa_{A Z\gamma}^2+ \kappa_{A Z\gamma}^{12} +\kappa_{A Z\gamma}^{21} $, where
\begin{align*}
&\kappa_{A Z \gamma }^i = 2 e  N_{c}^i Q_i \kappa_{ii} y_{ii}  \frac{M}{m_i} F_{1/2}^{(2)}(r_i, r_Z), \\
&\kappa_{A Z\gamma}^{ij} = 2 e N_c^i Q_{i} \kappa_{i j } y_{i j }   \left(\int_{0}^{1} dy \int_{0}^{1-y} dx \frac{4 M \left(\frac{r_{i}}{m_{i}} - (\frac{r_{i}}{m_{i}}- \frac {r_{j}}{m_{j}})x \right)}{r_{i}(1-x) + r_{j} x  + (r_{Z} -1) x y + r_{Z}(x^{2}-x)}\right).
 \end{align*}
The couplings $\kappa_{ij}$, $y_{ij}$ for each of the four cases are given in App.~\ref{effcoup_c1_append}, \ref{effcoup_c3_append}.
The expression for $F_{1/2}^{(2)}$ is a generalization to vector-like fermions of the expression given in Ref.~\cite{Gunion:1989we}. 

\medskip
\noindent\underline{$\kappa_{AZZ},~\kappa_{AWW}$}: 
Here we provide the expressions for $\kappa_{AZ Z}$ and $\kappa_{AWW}$ for $SVU$ and $SVQ$ models.
For the $SVU$ and $SVQ$ models,
\begin{align*}
\kappa_{AZZ}= 2  \left(\frac{g}{c_{W}}\right)^{2} \sum _i N_{c}^i (T^{i}_{3}- Q_{i}s_{W}^{2})^{2} y_{A}\frac{M}{m_i}  F_{1/2}^{(3)}(r_i, r_Z), \\
{\rm\text where} \ \
F_{1/2}^{(3)}(r_i, r_Z)= 4\int_{0}^{1}dy \int_{0}^{1-y} dx \frac{ r_i}{r_{i} - x y 
+ r_{Z} \left[  (x+y)^{2}-(x+y)\right]}.
\end{align*}
For $SVQ$ model,
$ \kappa_{A WW } = 2  \left(\frac{g}{\sqrt{2}} \right)^{2} \sum _i N_{c}^i y_{A}\frac{M}{m_i} F_{1/2}^{(3)}(r_i, r_W). $
For $SVU$~model $\kappa_{AWW}$ is zero.


\end{document}